\documentclass[12pt]{article} 

\usepackage{graphicx}
\usepackage{float}
\usepackage{color}
\usepackage{geometry}  
\usepackage{parskip}   
\usepackage{amsmath, amssymb} 
\usepackage{setspace} 
\usepackage{booktabs} 
\usepackage{enumerate}
\usepackage{dsfont}
\usepackage{tcolorbox}
\usepackage[framemethod=TikZ]{mdframed}
\usepackage{multicol}
\usepackage[export]{adjustbox}
\usepackage{url}
\usepackage{array}
\usepackage{multirow}
\usepackage{rotating}
\usepackage{bbm}
\usepackage{tabularray}
\usepackage[square,numbers]{natbib}
\begin{document}

\title{\vspace*{-2cm}Modeling the restricted mean survival time \\ using pseudo-value random forests}

\author{Alina Schenk$^{1,*}$, Vanessa Basten$^{1,2}$, Matthias Schmid$^{1}$}
\date{}

\maketitle

\begin{abstract}
\noindent The restricted mean survival time (RMST) has become a popular measure to summarize event times in longitudinal studies. Defined as the area under the survival function up to a time horizon $\tau > 0$, the RMST can be interpreted as the life expectancy within the time interval $[0, \tau ]$. In addition to its straightforward interpretation, the RMST also allows for the definition of valid estimands for the causal analysis of treatment contrasts in medical studies. In this work, we introduce a non-parametric approach to model the RMST conditional on a set of baseline variables (including, e.g., treatment variables and confounders). Our method is based on a direct modeling strategy for the RMST, using leave-one-out jackknife pseudo-values within a random forest regression framework.  In this way, it can be employed to obtain precise estimates of both patient-specific RMST values and confounder-adjusted treatment contrasts. Since our method \mbox{(termed ``pseudo-value random forest'',~PVRF)} is model-free, RMST estimates are not affected by restrictive assumptions like the proportional hazards assumption. Particularly, PVRF offers a high flexibility in detecting relevant covariate effects from higher-dimensional data, thereby expanding the range of existing pseudo-value modeling techniques for RMST estimation. We investigate the properties of our method using simulations and illustrate its use by an application to data from the SUCCESS-A breast cancer trial. Our numerical experiments demonstrate that PVRF yields accurate estimates of both patient-specific RMST values and RMST-based treatment contrasts.\\

\end{abstract}

\maketitle

\renewcommand\thefootnote{}
\footnotetext{$^1$Institute for Medical Biometry, Informatics and Epidemiology, University of Bonn, Venusberg-Campus~1, D-53127 Bonn, Germany $^2$Department of Mathematics, Informatics and Technology, Koblenz University of Applied Sciences, Rhein-Ahr-Campus, Remagen, Germany $^*$To whom correspondence should be addressed, e-mail: alina.schenk@imbie.uni-bonn.de}

\renewcommand\thefootnote{\fnsymbol{footnote}}
\setcounter{footnote}{1}

\section{Introduction}
\label{Introduction}

During the past years, an increasing number of statisticians and applied researchers has advocated the use of the restricted mean survival time (RMST) to summarize event times in longitudinal studies \cite{Royston2013, Ambrogi2022}. Defined as the area under the survival function within a time interval~$[0, \tau]$, the RMST represents the expected event time between zero and the ``time horizon'' $\tau > 0$. In medical research, using the RMST as a summary measure offers the following specific advantages: (i) its interpretation as the life expectancy between 0 and $\tau$ is straightforward and easily understood by both clinicians and patients \cite{McCaw2019}, (ii) instead of a single time point (evaluated, e.g., by $t$-year survival probabilities in cancer research), the entire survival history up to $\tau$ is reflected by the RMST, (iii) in contrast to the hazard ratio (HR) derived from Cox regression, the RMST can be used for meaningful treatment comparisons even when the proportional hazards (PH) assumption is violated \cite{Royston2013, Royston2011}, and (iv) the RMST can be used to define estimands for the causal interpretation of treatment and interventional effects \cite{Ni_2021}. As a result, the reporting of the RMST in medical studies has become increasingly prevalent \cite{Uno_2014, Dehbi_2017, stensrud2020}. 

In addition to the calculation of absolute RMST values, {\em differences} between group-wise RMST values have been suggested as a measure of treatment contrasts in longitudinal studies \cite{Uno_2014}. In medical research, treatment contrasts are often expressed and evaluated by the HR derived from a Cox proportional hazards model \cite{Royston2013}. However, the interpretation of this type of HR is only valid if the PH assumption holds; in particular, it assumes the HR to be constant over time. Thus, Stensrud and Hernán \cite{stensrud2020} recommended to supplement the reporting of HRs by summary measures directly derived from the survival \mbox{function $S(t) = \mbox{P}(T>t)$} (with $T$ denoting the survival time). The RMST belongs to this class of measures, as it can be expressed as \mbox{$\mu(\tau)~=\mbox{E}\left[\min(T, \tau)\right] = \int_0^{\tau} S(t)\,d t$} and therefore directly summarizes the survival function in~$[0, \tau]$. Similarly, the RMST difference for two treatment groups $A$ and $B$ with survival functions $S_A(t)$ and $S_B(t)$, defined by~$\mu_A(\tau) - \mu_B(\tau)$ \linebreak $= \int_0^{\tau} \left(S_A(t) - S_B(t)\right)dt$, can simply be interpreted as the difference in life expectancy or as a gain (or loss) in event-free survival time before $ \tau$ \cite{McCaw2019}. 

This paper is concerned with the estimation of individual RMSTs $\mu(\tau \vert X_i) =$ \linebreak $\int_0^{\tau} S(t \vert X_i)\,d t, \, i = 1, \ldots, n$, from a set of~$n$~independent individuals with possibly right-censored event times and covariate values $X_i = (X_{i}^{(1)}, \ldots, X_{i}^{(p)})^\top \in \mathbb{R}^p$. Note that, for ease of notation and without loss of generality, we assume all treatment and interventional variables to be included in $X_i$. Our method is characterized by a non-parametric approach combining pseudo-value modeling \cite{Anderden_RMST} with random forest regression \cite{breiman2001random, Mogensen_Gerds_2013} (for details, see below). Using the estimated individual RMSTs, we pursue two goals: (a) to incorporate the effects of a (possibly large and interacting) set of covariates in the estimation of the RMST, and (b) to quantify and assess accuracy of treatment effect estimation through RMST differences in observational longitudinal trials. 

Standard approaches to estimate individual RMSTs $\mu(\tau \vert X_i)$ are the direct integration of (group-wise) Kaplan-Meier curves (leading to identical RMST estimates for individuals belonging to the same treatment group) and the integration of survival functions estimated through a parametric or semi-parametric time-to-event model with covariates $X_i$ (e.g., a Cox PH model or an accelerated failure time (AFT) model \cite{Royston2011, Leurgans1987}). Using these standard approaches, the estimation of treatment effects through RMST differences is straightforward. Previous research on RMST differences also includes the work by Royston \& Parmar \cite{Royston2013}, Tian et~al. \cite{Tian_2018} and Huang \& Kuan \cite{Huang_Kuan_2018}, who developed hypothesis tests for RMST differences derived by group-wise integration of Kaplan-Meier curves. Clearly, the covariate-free Kaplan-Meier approach is not recommended for use in non-randomized studies, as it ignores the effects of potential confounders on RMST differences. While integrating estimated survival functions derived from Cox or AFT models mitigates this problem, the validity of the resulting RMST estimates (as well as their differences) strongly depends on the correctness of the underlying model and/or distributional assumptions \cite{Rong_2022}.

Instead of estimating individual RMSTs by integrating survival functions derived from time-to-event models, several authors have suggested to {\em directly} model the RMST \cite{Tian_2014, Wang_Schaubel_2018, Hasegawa_2020}. In general, the idea of direct modeling approaches is to estimate unconditional individual RMSTs (without using any covariate information) and to subsequently fit a statistical model regressing these values to the covariates. Key advantages of directly modeling the RMST are less restrictive distributional assumptions as well as the straightforward interpretation of the model coefficients \cite{Royston2013, McCaw2019, Uno_2014, Kim2017}. 

In this paper, we pursue a direct approach for modeling RMST values and their differences. More specifically, the idea of our method is to derive unconditional RMST values from jackknife pseudo-values and to regress these values to the covariates using random forests. Classical pseudo-value regression for the RMST difference \cite{Andersen_PoharPerme} is based on parametric models of the form
\begin{alignat}{3}
\label{eq:modellgleichung}
&g[\mu(\tau \vert X_i)] &&= \alpha + \gamma^T X_i &&=: \eta_i\, ,
\end{alignat}
with a monotonic link function $g$, an intercept $\alpha$ and covariate effects $\gamma$. Note that we suppress the dependency of $\alpha$, $\gamma$ and $\eta_i$ on $\tau$ for ease of notation.
Andersen et al.~\cite{Anderden_RMST} and Andersen \& Pohar Perme \cite{Andersen_PoharPerme} suggested to estimate unconditional RMST values by leave-one-out jackknife pseudo-values $\hat{\theta}_i(\tau)$ defined as
\begin{align}
\label{eq:PV}
\hat{\theta}_i(\tau) &\, = \, n \cdot \int\limits_0^{\tau} \hat{S}_{\text{KM}}(t)\,dt \, - \,
(n-1) \cdot \int\limits_0^{\tau} \hat{S}_{\text{KM}}^{-i}(t)\,dt\, , \, \ \ \ i=1,\ldots , n , 
\end{align}
where $\hat{S}_{\text{KM}}(t)$ denotes the Kaplan-Meier estimate evaluated at $t$ calculated on the complete data set and $\hat{S}^{-i}_{\text{KM}}(t)$ denotes the respective Kaplan-Meier estimate calculated on the data set without individual $i$. 

The coefficients in \eqref{eq:modellgleichung} can be estimated by a generalized estimation equation (GEE) approach, with $g$ being the identity or the log link \cite{Anderden_RMST}. However, while the GEE approach yields consistent estimates ($n \to \infty$) under the assumption of random \mbox{censoring} \cite{Graw_PseudoValues, Overgaard2017}, its flexibility is limited by the restrictive specification of the predictor $\eta_i$ in (\ref{eq:PV}). More specifically, $\eta_i = \alpha + \gamma^T X_i$ is defined by a linear combination of main effects terms, which might be too simplistic to describe the underlying data-generating process. Although more flexible effect terms (representing e.g.\@ interaction terms or non-linear main effects) could be included in (\ref{eq:modellgleichung}), this approach is not commonly used in practice. Often, this is due to the fact that pre-specifying an extended version of $\eta_i$ requires detailed knowledge on the, usually unknown, dependency structure between the pseudo-value outcome and the covariates. Further, the GEE approach does, in its basic form, neither incorporate any mechanism for data-driven variable selection nor perform any other sort of regularization to reduce redundant or irrelevant information. 

To address these issues, and to achieve the goals stated in (a) and (b), we propose to replace the established GEE approach by a random forest regression \cite{breiman2001random}. This regression model, which uses the pseudo-values $\hat{\theta}_i(\tau)$ in (\ref{eq:PV}) as continuous outcome and which will be termed ``pseudo-value random forest'' (PVRF) in the following, allows for an efficient data-driven selection of covariates and their interaction effects. In this way, the need to pre-specify $\eta_i$ is eliminated, making PVRF a particularly convenient method for applications involving a large number of covariates compared to the number of individuals (for instance, in medium-sized observational studies containing many potential confounders). By applying a g-computation formula \cite{Robins_1986, Snowden_2011} to the estimated RMST values, PVRF further allows for the direct estimation and causal interpretation of RMST differences. To additionally facilitate interpretability of the covariate effects, we propose to use state-of-the art methods for interpretable machine learning, as described in Molnar \cite{molnar2022}. 

The remainder of this paper is organized as follows: In Section \ref{pseudo_values_and_GEE}, we start with the definition and the properties of pseudo-values for RMST estimation, along with a brief description of the standard GEE modeling approach. Section~\ref{random_forest} provides a detailed description of the PVRF method, showing, in particular, how to derive individual estimates of the RMST. In Section~\ref{RMST_difference}, we describe how to apply g-computation for evaluating treatment contrasts by confounder-adjusted RMST differences. Sections~\ref{Experiments} and \ref{Results} contain the results of a simulation study investigating the properties of the PVRF method and comparing the proposed approach to established methods for RMST estimation. Section \ref{Application} presents an application of the PVRF method to data from the SUCCESS-A trial, a randomized phase III trial investigating the effects of two treatment regimens on the disease-free survival of patients with early breast cancer \cite{GregorioHaeberleFaschingetal.2020}. The main findings of the paper are summarized in Section \ref{Discussion}, along with a brief overview and discussion of related approaches. 

\section{Methods}
\label{Methods}

As stated above, we consider a set of $n$ independent individuals that are subject to right-censoring. For each \mbox{individual $i~=~1, \ldots, n$}, we consider covariate values $X_i = (X_{i}^{(1)}, \ldots, X_{i}^{(p)})$ measured at baseline. The individual survival time and censoring time are denoted by $T_i$ and $C_i$, respectively. The observed survival time is denoted by $\Tilde{T}_i = \min(T_i, C_i)$, and the status variable $\delta_i = \mathds{1}_{\lbrace{C_i > T_i\rbrace}}$ indicates whether the $i$-th individual is censored ($\delta_i = 0$) or whether the event of interest has been observed\linebreak  ($\delta_i = 1$). Following Graw et al.~\cite{Graw_PseudoValues}, we assume that the censoring times are independent of both the covariates and the event times.

\subsection{Estimation and modeling of the RMST via pseudo-values}
\label{pseudo_values_and_GEE}

When using the RMST, defined as $\mu(\tau) = \mbox{E}\left[\min(T, \tau)\right] = \int_0^{\tau} S(t)\,d t$, as dependent variable in a regression model, the outcome values are given by $\mu_i(\tau)$~$= \min(T_i, \tau)$, $i = 1,...,n$. By definition, these values depend on the survival times $T_i$, and, due to censoring, cannot be observed for all individuals. Pseudo-value regression overcomes this problem by replacing the partly incompletely observed outcome values with continuous (real-valued) pseudo-values $\hat{\theta}_i(\tau)$ that can be computed for both censored and uncensored individuals in the data set. For the RMST, the $i$-th pseudo-value at a time horizon $\tau$ is given by the right-hand side of Equation (\ref{eq:PV}). The pseudo-values can subsequently be used as a (completely observed) imputation for the outcome variable~$\mu_i(\tau)$ in the RMST regression model, facilitating the application of conventional modeling techniques like linear regression or decision trees~\cite{Schenk_2024}. Overgaard et al.\@ \cite{Overgaard2017} showed that the replacement of $\mu_i(\tau)$ with the pseudo-values~$\hat{\theta}_i(\tau)$ enables the consistent estimation of covariate effects on the RMST (see Overgaard et al.~\cite{Overgaard2017} for details and regularity assumptions).

\begin{figure}[h!]
\centering
\includegraphics[width=0.9\textwidth]{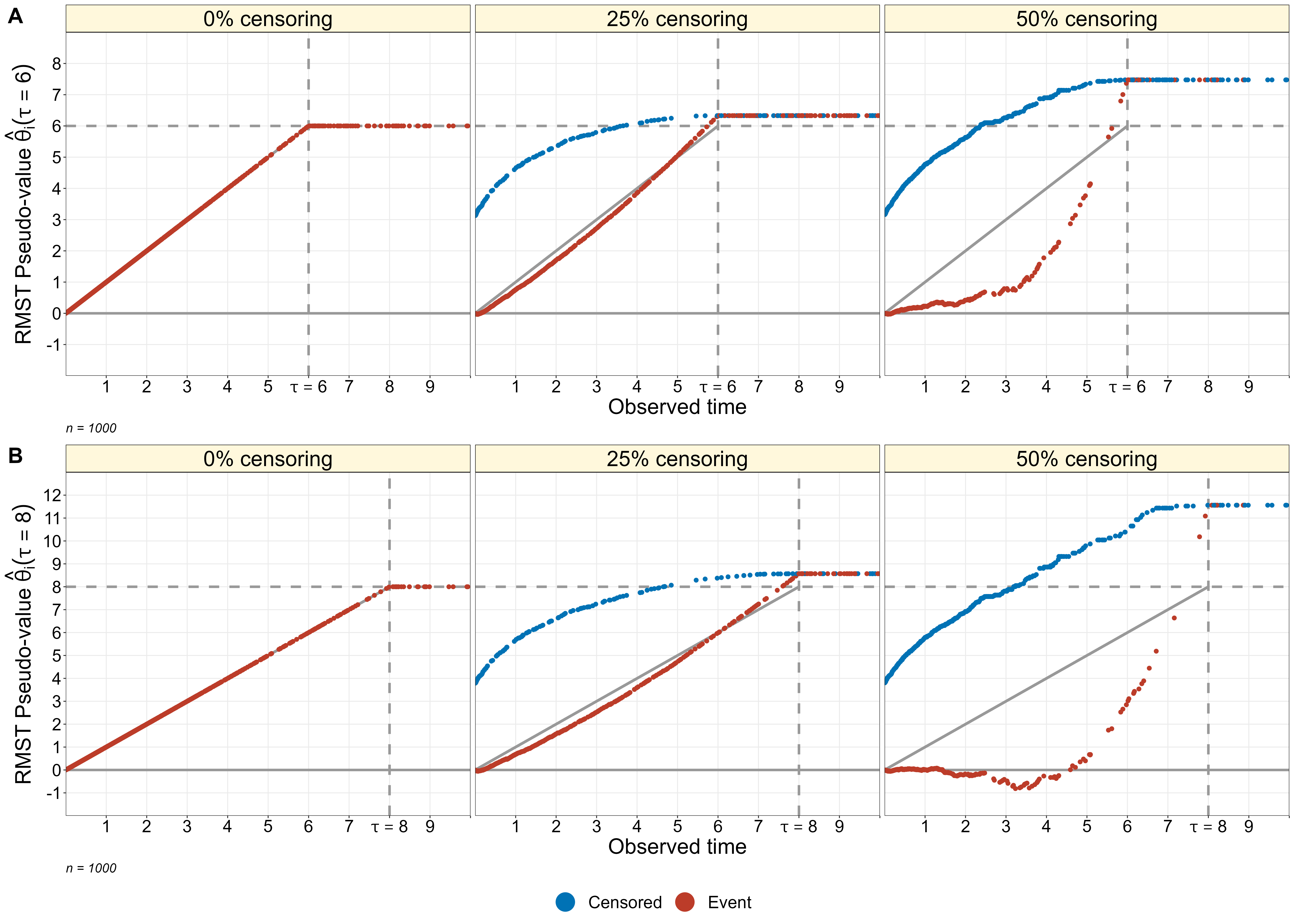}
\caption[Properties of pseudo-values]{Illustration of pseudo-values for the RMST, as derived from a data set with $n = 1000$. The time horizons were set to $\tau = 6$~(A) and $\tau = 8$~(B). Censored individuals are marked blue, individuals with an event are colored red. The dashed vertical lines indicate the time horizon $\tau$. 
In the data set with no censoring (left panels), the pseudo-values are identical to the observed event times for $\Tilde{T}_i < \tau$ and equal to $\tau$ for $\Tilde{T}_i \geq \tau$. For censored individuals in data sets with censoring (blue dots in the middle and right panels), it is observed that $\hat{\theta}_i(\tau) > \Tilde{T}_i$ for $\Tilde{T}_i < \tau$, irrespective of the choice of $\tau$ and the censoring proportion. For individuals with an observed event (red dots), there is no consistent pattern regarding the dependency of pseudo-values on~$\Tilde{T}_i$: In the middle panel of (A), referring to a smaller value of $\tau$ and a lower censoring proportion, pseudo-values of individuals with an observed event closely resemble the observed times. In contrast, in the right panel of (B), referring to a larger value of~$\tau$ and a higher censoring proportion, pseudo-values of individuals with an observed event are mostly lower than the observed event time and can even become negative. For $\Tilde{T}_i \geq \tau$, it holds that $\hat{\theta}_i(\tau) > \tau$ for all censored and uncensored individuals. The Figure has been adapted from Andersen \& Pohar Perme \cite{Andersen_PoharPerme}.} 
\label{fig:properties_pv}
\end{figure}

As seen from Figure \ref{fig:properties_pv}, the characteristics of pseudo-values for the RMST depend on the observed time $\Tilde{T}_i$, the censoring proportion in the data set and the time horizon $\tau$. If there is no censoring in the data (left panels in Figure \ref{fig:properties_pv}), it holds \mbox{that~$\hat{\theta}_i(\tau) = \Tilde{T}_i = T_i$} if $\Tilde{T}_i < \tau$ and $\hat{\theta}_i(\tau) = \tau$ if $\Tilde{T}_i \geq \tau$. For censored individuals (middle and right panels in Figure \ref{fig:properties_pv}), pseudo-values for the RMST are always larger than the observed survival time $\Tilde{T}_i$ (given that $\Tilde{T}_i < \tau$). On the other hand, pseudo-values of individuals with an observed event are mostly smaller than $\Tilde{T}_i$ and might even become negative when $\Tilde{T}_i < \tau$. For $\Tilde{T}_i \geq \tau$, it holds that $\hat{\theta}_i(\tau) > \tau$ for both censored and uncensored individuals. Consequently, there is no difference between individuals who were censored after time $\tau$ and those who were observed to experience an event after $\tau$ \cite{Andersen_PoharPerme}. In general, it appears hard to approximate the empirical distribution of the pseudo-values by a parametric distribution, as the former strongly depends on the time horizon~$\tau$, the distribution of the observed times, and the censoring pattern in the data. 

As outlined in Section \ref{Introduction}, the standard approach to pseudo-value regression for the RMST is to use the unconditional pseudo-values $\hat{\theta}_i(\tau)$ as outcome variable in a GEE model. For estimating the covariate effects using the GEE approach, it is convenient to re-write Equation \eqref{eq:modellgleichung} as
\begin{align}
\label{eq:GEE_approach}
g\left[\mu(\tau \vert X_i)\right] &= \alpha + \gamma^T X_i = \beta^T X_{i,.} = \eta_i \, ,
\end{align}
where the augmented covariate vector $X_{i,.} = (1, X_i^\top)^\top \in \mathbb{R}^{p+1}$ contains a dummy variable for the intercept $\alpha$. When modeling the RMST, $g$ is typically set to the identity link or the log link. The GEE estimate of $\beta \in \mathbb{R}^{p+1}$ is given by the solution to
\begin{align}
\label{eq:GEE}
\sum\limits_i \Big\{\frac{\partial}{\partial \beta}\, g^{-1}(\beta^TX_{i,.}) \Big\}^T V_i^{-1} \Big\{\hat{\theta}_i(\tau) - g^{-1}(\beta^TX_{i,.})\Big\} = 0 \, ,
\end{align}
where $V_i \in \mathbb{R}^+$ defines a working variance. In practice, $V_i$ is often set to 1, as Klein~\& Andersen \cite{Klein_Anderson} found no advantage of using more complex variance terms. The estimated coefficients $\hat{\beta}$ derived by solving~\eqref{eq:GEE} can be interpreted as direct effects on the RMST if $g$ is the identity link, or on the logarithm of the RMST if $g$ is the log link. A sandwich estimator can be used to estimate the variance of $\hat{\beta}$ \cite{Graw_PseudoValues}.

Despite the popularity of the GEE approach, it is easily seen from \eqref{eq:GEE_approach} that the estimated values of the RMST are constrained to a rather restrictive linear combination of the covariate effects. Particularly, as stated in Section~\ref{Introduction}, $\eta_i$ does not include any interaction terms. While these terms can be incorporated by pre-specifying them in $\eta_i$, it is well known that the number of interactions grows exponentially with the number of covariates. This implies numerous coefficients to be estimated and makes the GEE approach numerically unstable even in cases with a moderate number of covariates. In particular, it results in a high variance of the GEE estimator. If there is no expert knowledge available to pre-select a suitable (small) number of interaction terms, data-driven variable selection techniques could be applied. However, classical variable selection algorithms (such as forward, backward, or stepwise selection) also show a high variability when including a larger number of interaction terms. As a solution to the issues described above, we propose to apply random forest techniques for modeling the RMST. 

\subsection{Random forest regression}
\label{random_forest}

To address the issues described in Section \ref{pseudo_values_and_GEE}, we propose to model individual RMSTs\linebreak  by using the pseudo-values $\hat{\theta}_i(\tau)$ as continuous outcome variable in a random forest regression model \cite{breiman2001random, Mogensen_Gerds_2013}. Random forest regression is characterized by averaging estimates of multiple regression trees trained on different random subsets of the data (``ensemble'' of regression trees). In this way, overfitting is avoided, and both interactions and non-linear covariate effects can be captured by the model~\cite{breiman2001random}.
The general idea of building a regression tree is to derive local estimates of the outcome variable by partitioning the covariate space into a set of mutually exclusive subspaces \cite{breiman1984CART, Hothorn_conditional_inference_trees, treeBook}. Beginning with the \textit{root node} containing all individuals $i = 1, \ldots, n$, the idea is to successively evaluate a split criterion and to split individuals into two mutually exclusive sets termed \textit{daughter nodes}. Each daughter node $R_m \subseteq \lbrace{1, \ldots, n \rbrace}$ is further split into two daughter nodes $R_{m_1} \subset R_m$ and $R_{m_2} \subset R_m$ with $R_{m_1} \cap R_{m_2} = \emptyset$, and splitting is continued until some stopping criterion applies (see Appendix~B). In each node $R_m$, splitting is done by selecting a split variable $X^{(j)}$ and a corresponding split rule $\mathcal{R}_{mj}$ that optimize a pre-defined split criterion (e.g.~the mean squared error). The split criterion is evaluated on the data of the individuals in the respective node. In case of a continuous split variable $X^{(j)}$, the split rule $\mathcal{R}_{mj}$ is given as ``$X_i^{(j)} \leq \xi_{mj}$ vs.~$X_i^{(j)} > \xi_{mj}$" defining two mutually exclusive subsets $A(\mathcal{R}_{mj})$~$=\lbrace{X_i^{(j)} \vert \, i \in R_m \text{ and } X_i^{(j)} \leq \xi_{mj}\rbrace}$ and $A^*(\mathcal{R}_{mj})$~$=\lbrace{X_i^{(j)} \vert \, i \in R_m \text{ and } X_i^{(j)} > \xi_{mj}\rbrace}$, where $\xi_{mj} \in \mathbb{R}$ is the threshold of the split rule $\mathcal{R}_{mj}$ selected in node $R_m$. In case of a categorical split variable $X^{(j)}$, the split rule $\mathcal{R}_{mj}$ defines two mutually exclusive subsets $A(\mathcal{R}_{mj})$ and $A^*(\mathcal{R}_{mj})$ of the categories of $X^{(j)}$. In either case, the daughter nodes $R_{m_1}$ and $R_{m_2}$ are given as 
\begin{align}
R_{m_1} &= \lbrace{i \, \vert \, X_i^{(j)} \in A(\mathcal{R}_{mj}) \rbrace} \, , \notag \\
R_{m_2} &= \lbrace{i \, \vert \, X_i^{(j)} \in A^*(\mathcal{R}_{mj}) \rbrace} \, .
\end{align}
Nodes that are not further split into two daughter nodes because the stopping criterion applies are referred to as \textit{leaf nodes}. The route from the root node to a leaf node is termed \textit{path}. For calculating the estimated RMST value of a single individual, the associated leaf node is determined by successively applying the split rules from the root node to the leaf node. Afterwards, the RMST value is estimated by averaging the observed pseudo-values $\hat{\theta}_i(\tau)$ in the leaf node. 

Random forest regression is usually characterized by growing large ensembles of regression trees. In this paper, we will use~500 trees unless stated otherwise. Furthermore, we follow the strategy recommended by de Bin et al.\cite{Bin.2016} and grow our tree ensemble on subsamples of the complete data without replacement. Thus, each tree in the forest is grown on a different subset of the data, leading to different split rules and different RMST estimates in the leaf nodes. Additionally, only a random subset of the covariates is considered for splitting the nodes of the regression trees. In practice, the size of this subset (termed ``mtry'') is either set to $\sqrt{p}$ \cite{ranger_R_package} or optimized using cross-validation. In this paper, we applied five-fold cross-validation to determine the value of mtry, see Appendix~B. The final RMST estimate for individual~$i$ was obtained by dropping the covariate values $X_i$ down to the leaf nodes of the 500 trees and by averaging the 500 tree estimates. 

In the literature, there exist multiple tree building algorithms that vary in the procedure to select the split variables and the corresponding split rules. In this work, we consider two different tree building algorithms that will be described briefly in Sections \ref{subsub:CART_random_forest} and \ref{subsub:Conditional_random_forest}: (i) classification and regression trees (CART) \cite{breiman1984CART} and (ii) conditional inference trees \cite{Hothorn_conditional_inference_trees}. Correspondingly, the resulting forests will be referred to as \textit{CART random forest} and \textit{conditional random forest}.

\subsubsection{CART Random Forest}
\label{subsub:CART_random_forest}

In each node $R_m$, the CART algorithm selects the split variable $X^{(j)}$ and the corresponding split rule $\mathcal{R}_{mj}$ by minimizing the criterion
\begin{align}
\label{eq:CART_criterion}
\mbox{MSE}_{\mathcal{R}_{mj}} = \sum_{i \in R_{m_1}} \left(\hat{\theta}_i(\tau) - \overline{c}_1\right)^2 +
\sum_{i \in R_{m_2}} \left(\hat{\theta}_i(\tau) - \overline{c}_2\right)^2
\end{align}
over $\mathcal{R}_{mj}$, where $\overline{c}_1$ and $\overline{c}_2$ are the averaged pseudo-values in the daughter nodes $R_{m_1}$ and~$R_{m_2}$, respectively. Consequently, the split variable and the split rule minimizing the mean squared errors of the pseudo-values $\hat{\theta}_i(\tau)$ in the daughter nodes are selected jointly by one minimization step. In practice, this leads the CART algorithm to favor split variables with many possible splits, implying that the algorithm is usually biased towards the selection of covariates with many possible splits (e.g.\@ continuous covariates)~\cite{Hothorn_conditional_inference_trees}.

\subsubsection{Conditional Random Forest}
\label{subsub:Conditional_random_forest}

Unlike the CART algorithm, conditional inference trees \cite{Hothorn_conditional_inference_trees} follow a two-step process, selecting the optimal split variable by a set of statistical hypothesis tests {\it before} determining the corresponding split rule. In this way, a selection bias towards covariates with many possible splits is avoided. More specifically, in the first step, given a node~$R_m$ with \mbox{data $\mathcal{L}_m = \lbrace (\hat{\theta}_{i}(\tau), X_{i}^{(1)},\ldots, X_{i}^{(p)}) \, \vert\, i \in R_m \rbrace$},
the split variable is selected as the covariate showing the strongest association with the outcome variable~$\hat{\theta}_{i}(\tau)$. Associations are measured by the generalized correlation coefficient
\begin{align}
\label{eq:conditional_inference_T}
\rho_j(\mathcal{L}_m) = \mbox{vec}\left( \sum_{i \in R_m} \Tilde{g}_j (X_{i}^{(j)}) \cdot \hat{\theta}_i (\tau)\right) \in \mathbb{R}^{\Tilde{p}_j} \, , \ \ \ j=1,\ldots , p \, ,
\end{align}
where $\Tilde{g}_j (\cdot ) \in \mathbb{R}^{\tilde{p}_j}$, $j=1,\ldots , p$, is a set of transformation functions depending on the measurement scales of the covariates. For our purposes, we set $\Tilde{g}_j (X_{i}^{(j)}) = X_{i}^{(j)}$ if the $j$-th covariate is measured on a continuous scale. For unordered and ordered factors, the functions $\Tilde{g}_j (X_{i}^{(j)})$ are given by a set of dummy variables with $\tilde{p}_j$ equal to the number of categories of $X^{(j)}$. For the hypothesis tests, the elements of~$\rho_j(\mathcal{L}_m)$ are first standardized (assuming independence between the covariates and the pseudo-values in node $m$) and then transformed using the absolute value function \cite{Hothorn_conditional_inference_trees}. The so-obtained values can be interpreted as the absolute correlations between the covariates $X^{(j)},\, j = 1, \ldots, p$, and the pseudo-values $\hat{\theta}_i(\tau)$ in node~$R_m$. Afterwards, the maximum values of the resulting standardized vectors are used to test the null hypotheses of independence between the covariates and the outcome. Altogether, there are $p$ maximum values, resulting in $p$ hypothesis tests. Using random permutations of the pseudo-values, the conditional distributions of the maximum values under the null (i.e., independence between the covariates and the pseudo-values) are derived. Finally, the covariate with minimum p-value in the permutation tests is selected as split variable. Importantly, since the p-values do not depend on the scales of covariates, the selection procedure does not show any systematic preference of covariates with many possible splits.

The second step is to derive the split rule associated with the selected split variable. Analogous to the CART algorithm, each possible split rule leads to two mutually exclusive sets of individuals $R_{m_1}$ and $R_{m_2}$. To determine the optimal split rule, the idea is to maximize the criterion
\begin{align}
\mbox{MAE}_{\mathcal{R}_{m}} = \biggl| \frac{\sum_{i \in R_{m}} \mathbbm{1}_{ \{ i \in R_{m_1}\}} \cdot \hat{\theta}_{i} (\tau) - \mu_{m_1} }{\sigma_{m_1}}\biggr| 
\end{align} 
over all split rules, where $\mu_{m_1}$ and $\sigma_{m_1}$ denote the conditional means and standard deviations, respectively, \mbox{of $\sum_{i \in R_{m}} \mathds{1}_{ \{ i \in R_{m_1}\}} \cdot~\hat{\theta}_{i} (\tau)$} (calculated in the same way as in the standardization of $\rho_j(\mathcal{L}_m)$ above, cf.\@ Hothorn et al. \cite{Hothorn_conditional_inference_trees}). By definition,~$\mbox{MAE}_{\mathcal{R}_{m}}$ measures the association between node membership and the pseudo-values, ensuring that the outcome values of the individuals in the two daughter nodes become maximally dissimilar. 

\subsection{Evaluating RMST differences}
\label{RMST_difference}
In medical research, a common aim is to compare subgroups of the population with regard to their survival behavior. Usually, these subgroups are defined by an intervention (e.g.\@ treatment vs.\@ control, see Section \ref{Application}) or by the presence of a risk factor. Following Royston \& Parmar \cite{Royston2011}, Uno et al. \cite{Uno_2014} and Dehbi et al. \cite{Dehbi_2017}, we quantify differences in the survival behavior of population subgroups (in the following termed ``treatment contrasts'') using differences in RMST values. In randomized controlled trials (RCTs), which usually allow for ignoring all covariates except the intervention due to the randomization procedure, treatment contrasts can simply be estimated by the differences of the average RMST values in the relevant groups. Clearly, the calculation of treatment contrasts is less trivial when additional covariates have to be taken into account, particularly in non-randomized studies where the covariates usually take the roles of confounders. In these cases, we propose to apply \mbox{g-computation} to estimate treatment contrasts \cite{Robins_1986, Snowden_2011}. More specifically, denoting the treatment variable of individual $i$ by $X_i^{(\text{trt})}$ and the respective confounders by~$X_i^{(-\text{trt})}$, we propose to calculate individual RMST differences ($i=1,\ldots , n$) as
\begin{align}
\hat{\Delta}_i(\tau) = \hat{\mu}(\tau \vert X_i^{(-\text{trt})}, X_i^{(\text{trt})} = A) - \hat{\mu}(\tau \vert X_i^{(-\text{trt})}, X_i^{(\text{trt})} = B) \, ,
\end{align}
where $\hat{\mu}$ denotes the RMST estimate obtained from the random forest model (see Hu et al. \cite{Hu_2021} for alternative ways to define and estimate survival treatment effects). Based on the individual RMST differences, the average treatment effect (= treatment contrast) is estimated by
\begin{align}
\label{eq:Delta_tau}
\hat{\Delta}(\tau) = \frac{1}{n} \sum_{i=1}^n \hat{\Delta}_i(\tau) =\frac{1}{n} \sum_{i=1}^n \hat{\mu}(\tau \vert X_i^{(-\text{trt})}, X_i^{(\text{trt})} = A) - \hat{\mu}(\tau \vert X_i^{(-\text{trt})}, X_i^{(\text{trt})} = B) \, . 
\end{align}

\subsection{Pseudo-value random forest}
\label{random_forest_pseudo_value_regression}

Summarizing Sections \ref{pseudo_values_and_GEE} to \ref{RMST_difference}, we define our proposed method (termed ``pseudo-value random forest'', PVRF) by the following steps: 
\begin{enumerate}
\item Calculate pseudo-values $\hat{\theta}_i(\tau),\, i = 1, \ldots, n$, for the RMST (Equation \eqref{eq:PV}).
\item Grow a random forest using either the CART algorithm (Section \ref{subsub:CART_random_forest}) or conditional inference trees (Section \ref{subsub:Conditional_random_forest}).
\item Estimate individual RMST values $\hat{\mu}(\tau \vert X_i),\, i = 1, \ldots, n$, from the fitted random forest.
\item Depending on the research question, 
\begin{enumerate}
\item proceed analyzing estimated individual RMST values using interpretable machine learning techniques (for details, see Section \ref{Application}). 
\item estimate treatment contrasts $\hat{\Delta}(\tau)$ from individual RMST differences $\hat{\Delta}_i(\tau)$, $i = 1, \ldots, n$ (Equation \eqref{eq:Delta_tau}). 
\end{enumerate}
\end{enumerate}

\section{Experiments}
\label{Experiments}
To investigate the performance of PVRF, we carried out a comprehensive simulation study in R (version 4.1.2 \cite{R_referenz}). The data-generating process was based on a time-to-event model with an additive combination of main and interaction effects. We analyzed the ability of the PVRF approach to estimate RMSTs and RSMT differences between treatment groups in the absence and presence of two-way interactions. To this end, we considered scenarios with time-constant and time-varying treatment effects. The simulation study was based on 100 Monte Carlo replications. In each replication, we generated a training data set for model building and a test data set for model evaluation, each of size $n = 1000$. 

Survival times $T_i$, $i = 1,\ldots, n$, were generated from a Weibull model with scale parameter $\lambda > 0$, shape parameter $\nu > 0$ 
and hazard function 
\begin{align}
h(t \vert X_i) = \lambda \cdot \exp\left(\eta_i(t)\right) \cdot \nu \cdot t^{\nu-1} \,,
\end{align}
where $\eta_i(t)$ is the (possibly time-dependent) linear predictor of individual $i$ (depending on $X_i$, see Equation \eqref{eq:eta}). The cumulative hazard function was given by
\begin{align}
H(t \vert X_i) = \lambda \cdot \exp\left(\eta_i(t)\right) \cdot t^{\nu}.
\end{align}
The censoring times were generated independently of the survival times, using the same Weibull model with $\eta_i(t) = 0$. The parameters $\lambda$ and $\nu$ were adjusted such that the data-generating process yielded the desired censoring proportions. 

Overall, we examined four scenarios, each differing in the calculation of $\eta_i(t)$. Each scenario was characterized by five continuous covariates (denoted by $X_i^{(j)}$, $j = 1, \ldots, 5$) and five dichotomous covariates (denoted by $X_i^{(j)}$, $j = 6, \ldots, 10$). The continuous covariates followed a multivariate normal distribution with zero mean and a covariance matrix as given in Table~\ref{tab:covariance_matrix} in Appendix~A. Dichotomous covariates were independent and followed a Bernoulli distribution with probability~0.5 each. In addition, we considered a dichotomous treatment variable $X_{i}^{(\text{trt})}$ (treatment A vs. treatment B, Bernoulli distributed with \mbox{probability~0.5}). Depending on the scenario, $X_{i}^{(\text{trt})}$ was either time-constant or time-varying (see Table~\ref{tab:scenarios}). The scenarios further differed in the structure of the interactions between the covariates and the strength of the treatment effects. In general, we considered predictors of the form
\begin{align}
\label{eq:eta}
\eta_i(t) = \sum_{j=1}^{10} \delta_j X_{i}^{(j)} + \sum_{\substack{l \in \lbrace{1, \ldots, 5\rbrace} \\ m \in \lbrace{1, \ldots, 5\rbrace}}\newline} \psi_{lm} X_{i}^{(l)} X_{i}^{(m)} + \sum_{\substack{r \in \lbrace{1, \ldots, 5\rbrace} \\ s \in \lbrace{6, \ldots, 10\rbrace}}\newline} \varphi_{rs} X_{i}^{(r)} X_{i}^{(s)} + \vartheta_{\text{trt}}(t) \mathbbm{1}_{\{X_{i}^{(\text{trt})} = B\}}\,,
\end{align}
where $\delta_j$, $j = 1, \ldots, 10$, denote main effects of the continuous and the dichotomous covariates, $\psi_{lm}$, $l, m \in \lbrace{1, \ldots, 5\rbrace}$, represent the interaction effects between the continuous covariates, $\varphi_{rs}$, $r \in \lbrace{1, \ldots, 5\rbrace}, s \in \lbrace{6, \ldots, 10\rbrace}$, represent the interaction effects between the continuous and the dichotomous covariates, and $\vartheta_{\text{trt}}(t)$ denotes the (possibly time-varying) dichotomous treatment effect. All main and interaction effects were sampled from a continuous uniform distribution on $[-1,1]$; they were generated independently of each other and were the same in all Monte Carlo replications. Furthermore, we added five independent standard normally distributed noise variables to the covariate set. These were independent of the other covariates and did not affect the predictor $\eta_i(t)$. In Scenarios 1 and 2, the treatment effect was time-constant, whereas in Scenarios 3 and 4, the treatment effect changed at the transition time $t_0$, resulting in crossing survival curves (Figure~\ref{fig:Visualization}). Scenarios 1 and 3 included only main effects, whereas Scenarios 2 and 4 additionally included interaction effects. Table~\ref{tab:scenarios} provides an overview of the four scenarios, and Figure~\ref{fig:Visualization} presents the group-wise Kaplan-Meier curves for each scenario. 

\begin{table}[t]
\centering
\scriptsize
\begin{tblr}{>{\centering}p{1.1cm}>{\centering}p{2cm}>{\centering}p{2cm}>{\centering}p{4.5cm}>{\centering}p{3.2cm}}
\SetCell[r=3]{} \textbf{Scenario} & \SetCell[r=3]{}\textbf{Effects of continuous covariates} & \SetCell[r=3]{}\textbf{Effects of dichotomous covariates} & \SetCell[r=3]{} \textbf{Interaction effects} & \SetCell[r=3]{} \textbf{Treatment effect} \\
& & & & \\
& & & & \\
\hline
\SetCell[r=2]{} 1 & $\delta_j \sim \mbox{U}(-1,1)$ & $\delta_j \sim \mbox{U}(-1,1)$ & $\psi_{lm} = 0 \, \forall l,m$ & \SetCell[r=2]{} $\vartheta_{\mbox{trt}}(t) = -2$ \\
& $j = 1, \ldots, 5$ & $j = 6, \ldots, 10$ & $\varphi_{rs} = 0 \, \forall r,s$ & \\
\hline 
\SetCell[r=2]{} 2 & $\delta_j \sim \mbox{U}(-1,1)$ & $\delta_j \sim \mbox{U}(-1,1)$ & $\psi_{13}, \psi_{14}, \psi_{23}, \psi_{25}, \psi_{45} \sim \mbox{U}(-1,1)$ & \SetCell[r=2]{} $\vartheta_{\mbox{trt}}(t) = -2$ \\
& $j = 1, \ldots, 5$ & $j = 6, \ldots, 10$ & $\varphi_{17}, \varphi_{28}, \varphi_{39} \sim \mbox{U}(-1,1)$ & \\
\hline
\SetCell[r=2]{} 3 & $\delta_j \sim \mbox{U}(-1,1)$ & $\delta_j \sim \mbox{U}(-1,1)$ & $\psi_{lm} = 0 \, \forall l,m$ & \SetCell[r=2]{} $\vartheta_{\mbox{trt}}(t) = \begin{cases} -2, & t \leq t_0 \\ \hphantom{-}2, & t > t_0 \end{cases}$ \\
& $j = 1, \ldots, 5$ & $j = 6, \ldots, 10$ & $\varphi_{rs} = 0 \, \forall r,s$ & \\
\hline 
\SetCell[r=2]{} 4 & $\delta_j \sim \mbox{U}(-1,1)$ & $\delta_j \sim \mbox{U}(-1,1)$ & $\psi_{13}, \psi_{14}, \psi_{23}, \psi_{25}, \psi_{45} \sim \mbox{U}(-1,1)$ & \SetCell[r=2]{} $\vartheta_{\mbox{trt}}(t) = \begin{cases} -2, & t \leq t_0 \\ \hphantom{-}2, & t > t_0 \end{cases}$ \\
& $j = 1, \ldots, 5$ & $j = 6, \ldots, 10$ & $\varphi_{17}, \varphi_{28}, \varphi_{39} \sim \mbox{U}(-1,1)$ & \\
\hline 
\end{tblr}
\caption{Overview of the four scenarios used in the simulation study, each characterized by five continuous covariates, five dichotomous covariates, eight interaction effects (Scenarios 2 and 4), and a time-constant (\mbox{Scenarios 1 and 2}) or time-varying (Scenarios 3 and 4) treatment effect. All interaction effects not contained in the fourth column were set to zero. }
\label{tab:scenarios}
\end{table}

In each of the four scenarios, we considered three different censoring proportions (25\%, 50\%, 75\%) and five different values of the time horizon $\tau$. The latter were determined by the 50\%, 60\%, 70\%, 80\% and 90\% quantiles of the observed times~$\tilde{T}_i$, $i = 1, \ldots, n$, denoted by $q_{50\%}$, $q_{60\%}$, $q_{70\%}$, $q_{80\%}$ and $q_{90\%}$, respectively. The values of~$\tau$, which were held fixed across the simulation runs, are given in Table \ref{tab:tau_values} in Appendix~A. The transition time $t_0$ in Scenarios 3 and 4 was set to~$q_{70\%}$. In total, each scenario examined 15 combinations of censoring proportions and time horizons $\tau$. For the values of the \mbox{coefficients $\delta_j$, $\psi_{lm}$ and $\varphi_{rs}$}, we refer to the R code at {\tt https://www.imbie.uni-bonn.de/cloud/index.php/s/6gmJQmayFAMJZHk}. 

To evaluate performance of RMST estimates, we considered the mean squared error defined by
\begin{align}
\label{eq:MSE}
\text{MSE} \,=\, \frac{1}{n} \,\sum_{i=1}^{n} \left(\hat{\mu}(\tau| X_i) - \mu(\tau| X_i)\right)^2 \, ,
\end{align}
where $\hat{\mu}(\tau| X_i)$ and $\mu(\tau| X_i)$ denote the estimated and the theoretical RMSTs, respectively, of individual $i$ at time horizon $\tau$. The root mean squared error (RMSE) is defined as the square root of \eqref{eq:MSE}. The theoretical RMST in \eqref{eq:MSE} is derived as 
\begin{align}
\label{eq:RMST_theoretical}
\mu(\tau \vert X_i) &= \int_{0}^{\tau} S(t\vert X_i)\, d t = \int_{0}^{\tau} \exp(-H(t \vert X_i))\,d t \notag \\ &= 
\begin{cases}
\int_0^{\tau} \exp(-H_1(t\vert X_i))\,d t\,, & \tau \leq t_0\,, \\[.2cm]
\int_0^{t_0} \exp(-H_1(t\vert X_i))\,d t + \int_{t_0}^{\tau} \exp(-H_1(t_0\vert X_i) \\ -H_2(t\vert X_i)+H_2(t_0\vert X_i))\,d t\,, & \tau > t_0\,, 
\end{cases} 
\end{align}
where $H_1(t \vert X_i)$ and $H_2(t \vert X_i)$ are the cumulative hazard functions before and after the transition point $t_0$, respectively. Note that in Scenarios 1 and 2, the hazards are constant over time and thus, $H_1(t \vert X_i) = H_2(t \vert X_i)$, resulting in $\mu(\tau \vert X_i) = \int_0^{\tau} \exp(-H_1(t\vert X_i))d t$ for both $\tau \leq t_0$ and $\tau > t_0$. Analogously, evaluated the accuracy of treatment effect estimates by calculating the (root) mean squared error of the treatment effect, defined as 
\begin{align}
\label{eq:MSE_delta}
\text{MSE}_{\Delta} \,=\, \frac{1}{n} \,\sum_{i=1}^{n} \left(\hat{\Delta}_i(\tau) - \Delta_i(\tau)\right)^2 \, ,
\end{align}
where $\hat{\Delta}_i(\tau)$ and $\Delta_i(\tau)$ denote the estimated and the theoretical individual treatment effects, respectively, of individual $i$ at time horizon $\tau$ (see Section \ref{RMST_difference}).
In addition to evaluating the estimation accuracy of the CART random forest and the conditional random forest approaches, we compared our proposed method to alternative modeling approaches. These were (i) a GEE pseudo-value model with identity link (\textit{GEE}), (ii) a GEE pseudo-value model with log link (\textit{GEE (log)}), (iii) a Cox proportional hazards model (\textit{Cox}), \mbox{(iv) a parametric AFT} model (based on log-transformed survival times and assuming \mbox{normally distributed errors}~\cite{therneau_grambsch_2000},~\textit{Lognormal}), \mbox{and (v)} a correctly specified Cox proportional hazards model (\textit{Reference}) serving as benchmark for the performance measures. For the modeling approaches described in (i)-(iv), we specified the main effects of all available continuous and dichotomous covariates (including the noise variables) but did not consider any interaction terms. The \textit{Reference} model was specified such that it corresponded to the true data-generating process. More specifically, it incorporated the \mbox{\textit{informative} (= non-zero)} main and interaction effects only (see Table \ref{tab:scenarios}). In particular, the \textit{Reference} model accounted for the time-dependent treatment effect in Scenarios 3 and 4. This was accomplished by specifying a time-varying stratification variable that enabled the Cox model to estimate a time-dependent treatment effect. 
Consequently, \textit{Reference} served as a lower benchmark in the RMSE and $\mbox{RMSE}_{\Delta}$ comparisons. For the \textit{Cox}, \textit{Lognormal} and \textit{Reference} models, which do not directly model the RMST, estimates of the RMST were derived through the integration of the estimated survival function. Further details on the implementation of the methods are given in Appendix~B.
We anticipate that the CART random forest and the conditional random forest approaches will outperform the \textit{Cox}, \textit{Lognormal}, \textit{GEE}, and \textit{GEE (log)} approaches in the scenarios with non-zero interaction terms (Scenarios 2 and 4). Additionally, due to the time-varying treatment effect, we expect the pseudo-value methods to outperform the \textit{Cox} model in Scenarios 3 and 4. Generally, we expect both the RMSE and $\mbox{RMSE}_{\Delta}$ values to increase with $\tau$, since the RMST also rises with $\tau$.

\section{Results}
\label{Results}
\begin{figure}[t]
\centering
\includegraphics[width=\textwidth]{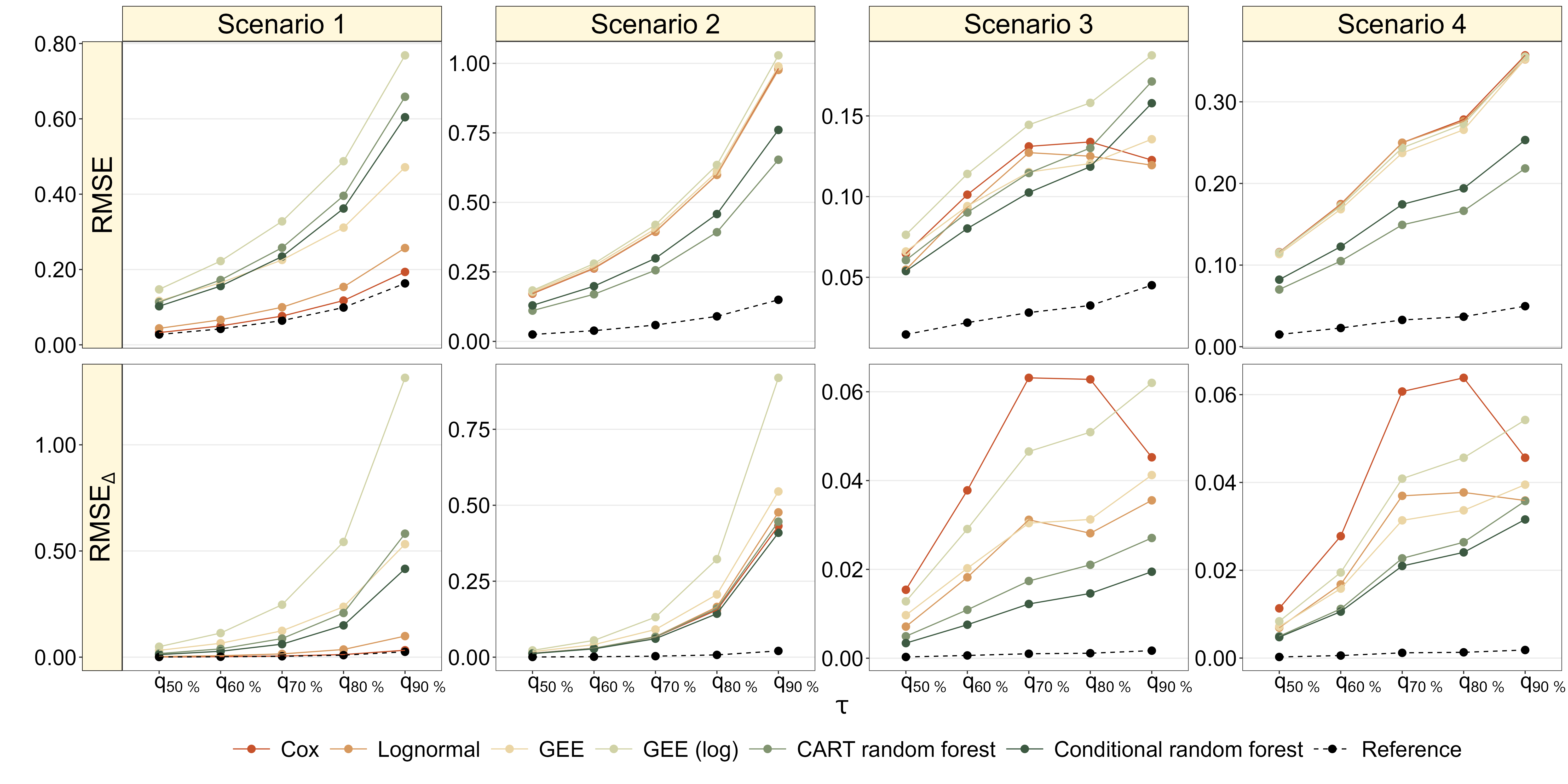}
\caption{Results of the simulation study (50\% censoring). The upper panels present the average RMSE \eqref{eq:MSE}, as obtained from the RMST estimates at different values of~$\tau$. The lower panels present the average RMSE for the treatment effect \eqref{eq:MSE_delta}. The green colored dots refer to the PVRF method (CART random forest and conditional random forest). The dashed black lines refer to the correctly specified Cox proportional hazards model (\textit{Reference}). } 
\label{fig:Results_Simulations}
\end{figure}

Figure \ref{fig:Results_Simulations} summarizes the simulation results of the four scenarios at a censoring proportion of 50\%. In the first scenario (main effects only, time-constant treatment effect), both the average RMSE and the average $\mbox{RMSE}_{\Delta}$ increase with $\tau$, as expected. This is true for all considered models. Notably, there is a clear difference in terms of RMSE between the standard modeling techniques (\textit{Cox} and \textit{Lognormal}) and the pseudo-value methods (\textit{GEE}, \textit{GEE (log)}, \mbox{CART random forest} and \mbox{conditional random forest}), with the best performing model being the \textit{Cox} model followed by the \textit{Lognormal} model. This result can be explained by the fact that the Cox model matches the data-generating mechanism (except for the noise variables). Among the pseudo-value regression methods, the conditional random forest demonstrates superior performance for $\tau\leq q_{60\%}$ followed by \textit{GEE}, \mbox{CART random forest} and \textit{GEE (log)}. However, this is no longer true when $\tau > q_{60\%}$. In terms of the RMSE for the treatment effect ($\mbox{RMSE}_{\Delta}$), the \textit{Cox} model demonstrates the best performance, in line with our expectations, followed by the \textit{Lognormal} model. Among the pseudo-value methods, the conditional random forest performs best with regard to treatment effect estimation, followed by the CART random forest. The application of the log link in the GEE approach (\textit{GEE (log)}) appears to have a negative effect on both performance measures (first column of Figure \ref{fig:Results_Simulations}).
In Scenario 2 (non-zero interaction effects, time-constant treatment effect), the average $\mbox{RMSE}_{\Delta}$ increases with $\tau$, similar to Scenario 1. The tree-based pseudo-value methods, particularly the CART random forest, performs best in terms of RMSE, having a slight advantage over the conditional random forest. This result demonstrates the ability of tree-based methods to identify and model interactions between the covariates. All other methods perform similar in this scenario. Regarding treatment effect estimation, the conditional random forest performed best, having slight advantages over the standard \textit{Cox} and \textit{Lognormal} modeling techniques, as well as over the CART random forest. Although the \textit{Cox} and \textit{Lognormal} models do not perform well in terms of RMSE, their performance regarding treatment effect estimation is comparable to the respective performance of the tree-based methods. On the other hand, the GEE with log link shows a poor performance in the estimation of the treatment effect, with estimates getting worse as $\tau$ increases (second column of Figure~\ref{fig:Results_Simulations}).

\begin{figure}[t]
\centering
\includegraphics[width=\textwidth]{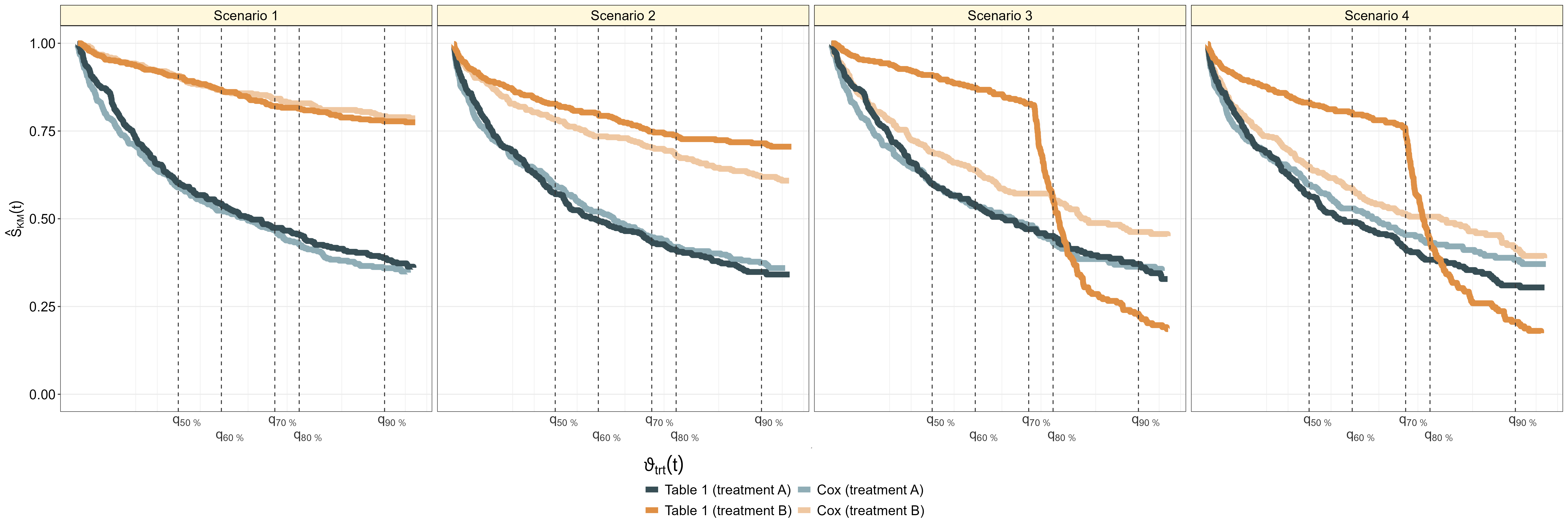}
\caption{Results of the simulation study (50\% censoring). The dark lines depict the Kaplan-Meier curves in the two treatment groups, as obtained from $n = 1000$ individuals with data generated according to Table~\ref{tab:scenarios} (including the true treatment effects $\vartheta_{\text{trt}}(t)$). The bright lines depict the Kaplan-Meier curves derived from data generated according to Table~\ref{tab:scenarios} but including the (time-constant) average treatment effect estimated by the \textit{Cox} method instead of the true treatment effect (Scenario~1: $\hat{\vartheta}_{\text{trt}}^{\text{Cox}}(t) = -2.10$, Scenario~2: $\hat{\vartheta}_{\text{trt}}^{\text{Cox}}(t) = -1.26$, Scenario~3: $\hat{\vartheta}_{\text{trt}}^{\text{Cox}}(t) = -0.38$, Scenario~4: $\hat{\vartheta}_{\text{trt}}^{\text{Cox}}(t) = -0.33$).}
\label{fig:Visualization}
\end{figure}

In Scenario 3 (main effects only, time-dependent treatment effect), the average RMSE values of the standard modeling techniques \textit{Cox} and \textit{Lognormal} do not monotonically increase with $\tau$, as in Scenarios 1 and 2. Instead, there is a turning point in the RMSE values at $\tau = t_0 = q_{70\%}$. This can be explained as follows: As seen from Figure~\ref{fig:Visualization}, the \textit{Cox} model assumes a time-constant treatment effect, implying that the effect of treatment $B$ is underestimated when~$t \leq t_0 = q_{70\%}$. This in turn leads to strongly biased RMST estimates. On the other hand, the \textit{Cox} model overestimates the effect of treatment $B$ when~$t > t_0 = q_{70\%}$. Consequently, as RMST estimates are derived by the area under the survival curve up to $\tau$, the part of the area under the survival curve that is not included in the RMST estimates for $t \leq t_0 = q_{70\%}$ is compensated by the excess area under the estimated survival curve for $t < t_0 = q_{70\%}$. As a result, the \textit{Cox} model yields decreasing RMSE values for~$\tau > t_0 = q_{70\%}$. For the \mbox{\textit{Lognormal}} model, analogous observations can be made. In contrast, the pseudo-value methods exhibit increasing RMSE values with rising~$\tau$, as expected. In general, the pseudo-value methods (except for \textit{GEE (log)}) outperform the \textit{Cox} and \textit{Lognormal} models with respect to the RMSE for $\tau \leq t_0$. Specifically, the conditional random forest demonstrates superior performance compared to the \textit{Cox} and \textit{Lognormal} models. On the other hand, as the RMSE values obtained from the \textit{Cox} and \textit{Lognormal} models decrease for $\tau > t_0$, these methods perform better than the pseudo-value methods at $\tau = q_{90\%}$. Regarding treatment effect estimation, the $\mbox{RMSE}_{\Delta}$ values obtained from the pseudo-value methods increase with $\tau$. In contrast, the $\mbox{RMSE}_{\Delta}$ values obtained from the \textit{Cox} model increase for $\tau \leq t_0 = q_{70\%}$ but decrease for $\tau > t_0 = q_{70\%}$. Again, this is due to the underestimation (overestimation) of the effect of treatment~$B$ for $t \leq t_0 = q_{70\%}$ ($t > t_0 = q_{70\%}$). The tree-based methods (CART random forest and conditional random forest) consistently perform best, regardless of the value of~$\tau$, confirming their ability to capture the time-dependent treatment effect in this scenario (third column of Figure \ref{fig:Results_Simulations}). 
In the presence of interaction effects, as in Scenario 4 (non-zero interaction effects, time-dependent treatment effect), there is a clear advantage of the tree-based methods (CART random forest and conditional random forest) with respect to both, $\mbox{RMSE}$ and $\mbox{RMSE}_{\Delta}$ (fourth column of Figure \ref{fig:Results_Simulations}). Again, these results highlight the ability of tree-based methods to model interaction effects and to capture time-dependent treatment effects. As seen from Figure \ref{fig:Results_Simulations}, the time-dependent treatment effect influences the trend of the $\mbox{RMSE}_{\Delta}$ values of the \textit{Cox} and \textit{Lognormal} models, similar to Scenario 3, but not the trend of the respective RMSE values. Of note, the CART random forest and conditional random forest methods show the best performance across all time horizons $\tau$. 

In summary, when considering scenarios with interactions and/or time-varying treatment effects, the proposed PVRF method (represented by the CART random forest and the conditional random forest) performs better than standard modeling approaches in estimating the RMST. The CART random forest has a slight advantage over the conditional random forest with respect to the RMSE in the presence of interaction effects, as observed in Scenarios 2 and 4. Regarding the estimation of treatment contrasts, the conditional random forest method performed best in the scenarios with interactions and/or time-varying treatment effects. The value of $\tau$ relative to the transition time $t_0$ significantly impacts the performance of the standard modeling techniques \textit{Cox} and \textit{Lognormal}, especially in the scenarios with time-varying treatment effects.
The results obtained for $25\%$ and $75\%$ censoring are similar to the results shown in Figures \ref{fig:Results_Simulations} and \ref{fig:Visualization}. They are presented in Figures \ref{fig:Results_Simulations_25} and~\ref{fig:Results_Simulations_75} in Appendix~C.

\section{Application}
\label{Application}
To illustrate the use of the PVRF approach, we applied the CART random forest and the conditional random forest methods to data from the multicenter randomized phase III SUCCESS-A trial (NCT02181101). SUCCESS-A enrolled 3,754 patients with a primary invasive breast cancer between September 2005 and March 2007~\cite{GregorioHaeberleFaschingetal.2020}. Study participants were randomly assigned in a~1:1 ratio to one of two treatment arms, which received either standard chemotherapy (control group) or standard chemotherapy with the addition of gemcitabine (interventional group). For details on the inclusion/exclusion criteria and the design of the study see de Gregorio et al.~\cite{GregorioHaeberleFaschingetal.2020}. 
The primary aim of the SUCCESS-A trial was to compare the two treatment arms with respect to disease-free survival (DFS), defined as the period from the date of randomization to the earliest date of disease progression (distant metastases, local and contralocal recurrence, and secondary primary tumors) or death from any cause \cite{GregorioHaeberleFaschingetal.2020, STEEP}. Here, we present the results of a secondary analysis that considered DFS as the outcome of interest. Note that the definition of DFS includes death from any cause. Accordingly, we did not consider death as a competing event.
Patients were censored at the last date at which they were known to be disease-free, resulting in an event proportion of $12.2\%$ (458 events in 3,754 patients). The maximum observation time was 5.5 years (6 months of chemotherapy followed by 5 years of follow-up; median 5.2 years, first quartile 3.7 years, third quartile 5.5 years). Patient characteristics included age at randomization ({\em age}, measured in years), body mass index ({\em BMI}, measured in $kg/m^2$) and menopausal status ({\em meno}, two categories, pre-/post-menopausal) as well as information on the tumor, including stage ({\em stage}, four \mbox{categories, pT1/pT2/pT3/pT4}), grade ({\em grade}, three categories, G1/G2/G3), lymph node status ({\em nodal status}, two \mbox{categories, pN0/pN+}), type ({\em type}, three categories, ductal/lobular/other) and receptor status of estrogen ({\em ER}), progesterone ({\em PR}), and {\em HER2} (two categories each, negative/positive). A descriptive summary of the variables is given in Table~\ref{tab:supplement_SUCCESS_A} in Appendix~D. Patients with missing values in any of the considered covariates were excluded from our anaylsis. The analyzed data \mbox{comprised 3,652} patients. 
The main aim of our analysis was to model the RMST for DFS at $\tau = 5$ years, corresponding to the length of the follow-up period. To this end, we applied the CART random forest, the conditional random forest, the \textit{Cox} model and the \textit{GEE} model to the SUCCESS-A study data. The covariates were defined by the treatment (control/intervention) and the ten patient \& tumor characteristics described above. Predictive accuracy of the models was measured by the weighted residual sum of squares~(WRSS), an inverse-probability-of-censoring (IPC) weighted error measure introduced by Cwiling et al. \cite{WRSS}, using five-fold cross validation. Denoting the data in the five test folds by $D_k \subset \lbrace{1, \ldots, n\rbrace}, k = 1, \ldots, 5$, the cross-validated WRSS is defined by
\begin{align}
\label{eq:WRSS}
\mbox{WRSS} = \frac{1}{5}\sum_{k=1}^5 \mbox{WRSS}_k = \frac{1}{5} \sum_{k=1}^5 \sum_{i \in D_k} \frac{1}{\vert D_k \vert} \left(\min(\Tilde{T}_i,\tau) - \hat{\mu}(\tau|X_{i})\right)^2 \cdot \hat{w}_i\, ,
\end{align}
with $\hat{\mu}(\tau|X_{i})$ denoting the estimated RMST for individual $i$ derived on the data in the training folds $\lbrace{1, \ldots, n\rbrace} \setminus D_k$. The IPC weights $\hat{w}_i$ are defined by
\begin{align}
\label{eq:weights_WRSS}
\hat{w}_i = \frac{\mathbbm{1}\{\Tilde{T}_i \leq \tau\} \cdot \delta_i}{\hat{G}(\Tilde{T}_{i^-}|X_{i})} + \frac{\mathbbm{1}\{\Tilde{T}_i > \tau\}}{\hat{G}(\tau|X_{i})} \, ,
\end{align}
where $\hat{G}$ is a consistent estimator of the censoring survival function. In this work, we use the Kaplan-Meier method to calculate~$\hat{G}$. The cross-validated treatment effect (measured in days, control vs.\@ interventional group) was calculated as 
\begin{eqnarray}~
\label{eq:estimated_delta_application}
 \overline{\hat{\Delta}}(\tau)  &=& \ \frac{1}{5} \sum_{k=1}^5 \sum_{i\in D_k} \frac{1}{\vert D_k \vert} \cdot \hat{\Delta}_i(\tau)  \nonumber\\
& = & \frac{1}{5} \sum_{k=1}^5 \sum_{i\in D_k} \frac{1}{\vert D_k \vert} \cdot \left[\hat{\mu}(\tau \vert X_i^{(-trt)}, X_{i}^{(trt)} = \text{control}) \right. \nonumber\\
&& \left. - \ \hat{\mu}(\tau \vert X_i^{(-trt)}, X_{i}^{(trt)} = \text{interventional}) \right].
\end{eqnarray} 
To enhance the interpretability of the random forest methods, we computed the permutation feature importance ($\mbox{PFI}_j$) and Shapley values for each covariate ($j=1,\ldots , 11$) \cite{molnar2022}. $\mbox{PFI}_j$ was determined on each test fold: Given the fitted model (here, either the CART random forest or the conditional random forest), $\mbox{PFI}_j$ is defined as the ratio of $\mbox{WRSS}_k$ with $\hat{\mu}(\tau \mid X_i)$ derived by the fitted model but with permuted values of the $j$-th covariate (numerator) and $\mbox{WRSS}_k$ with $\hat{\mu}(\tau \mid X_i)$ calculated as usual (denominator, see Equation \eqref{eq:PFI}). Thus, higher $\mbox{PFI}_j$ values indicate a higher importance of the $j$-th covariate for estimating the RMST. Local Shapley values were derived for 1000 randomly selected patients. A high absolute local Shapley value indicates a high importance of the respective covariate in the estimation of the RMST.

\begin{figure}[h!]
\centering
\includegraphics[width=0.87\textwidth]{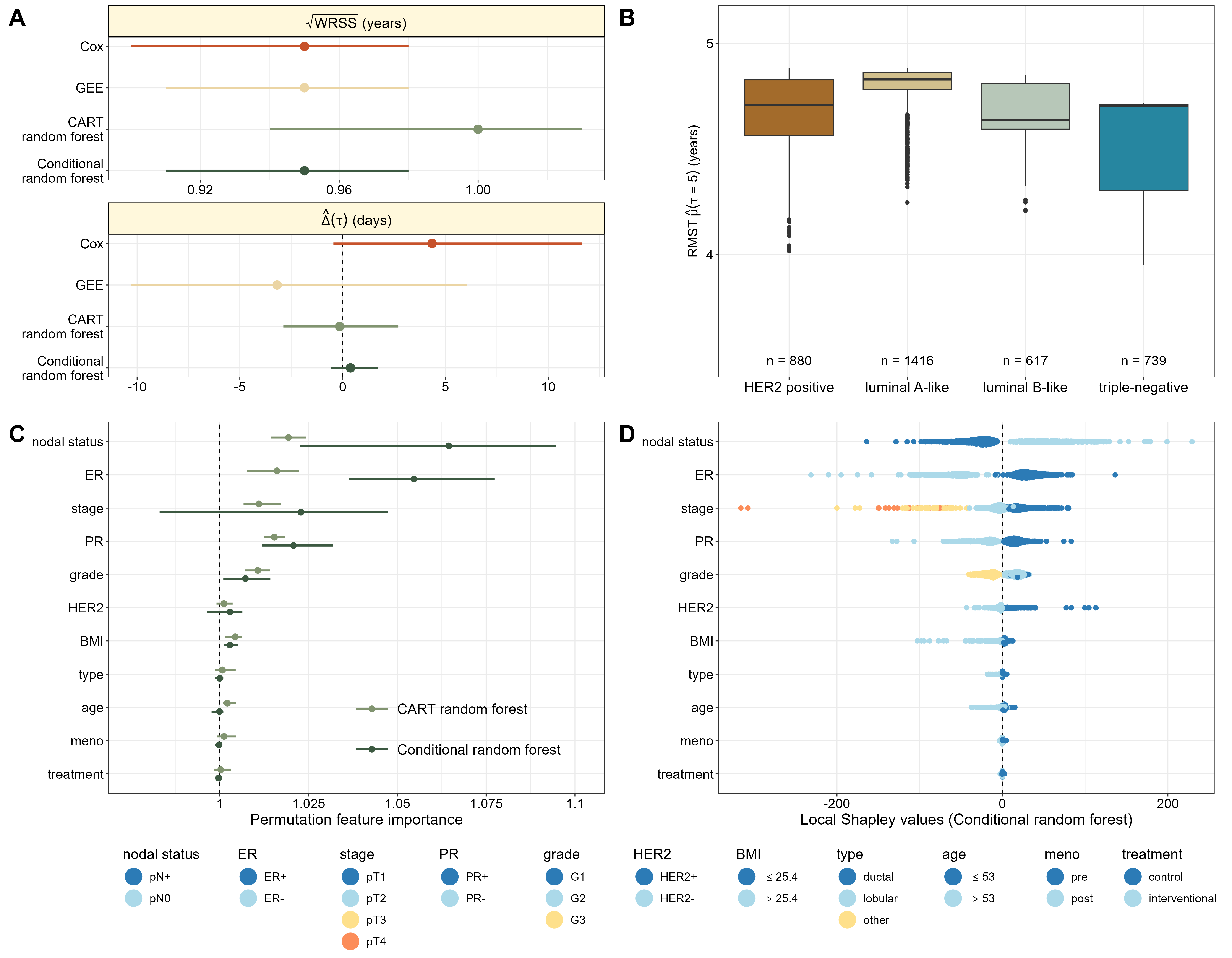}
\caption{Analysis of DFS in the SUCCESS-A study. The four panels present the results obtained from the \textit{Cox}, \textit{GEE}, CART random forest and conditional random forest methods. Panel (A) shows the WRSS (measured in years) and the treatment effect $\hat{\Delta}(\tau)$ (measured in days). The dots represent the five-fold cross-validated values of $\mbox{WRSS}$ and $\hat{\Delta}(\tau)$, while the bars refer to the respective ranges of $\mbox{WRSS}$ and $\hat{\Delta}(\tau)$ in the five folds. Panel (B) shows the estimated RMST values in patient groups defined by molecular tumor subtypes, as obtained from the conditional random forest method (see Table \ref{tab:subgroups_SUCCESS_A}). Panel (C) shows the permutation feature importance of each covariate for the CART random forest and the conditional random forest methods. The dots represent the five-fold cross-validated values of $\mbox{PFI}_j$ and the bars refer to the respective ranges in the five folds. Panel~(D) presents local Shapley values for each covariate, as obtained by evaluating the conditional random forest estimates in 1000 randomly selected patients. Each dot corresponds to one patient. The color codings used in Panel~(D) are presented in the bottom row of the figure. The estimated RMST values and local Shapley values obtained from the CART random forest are shown in Figure \ref{fig:SUCCESS_A_CART} in Appendix~D.} 
\label{fig:Results_SUCCESS_A}
\end{figure}
The results of our analysis are presented in Figure \ref{fig:Results_SUCCESS_A}. They show that the two random forest methods detected several established prognostic factors and subgroups, which have been consistently reported in the literature and have also entered treatment guidelines for primary breast cancer \cite{Coates_2015, Senkus_2015}. Regarding the WRSS (Figure \ref{fig:Results_SUCCESS_A} (A)), the conditional random forest with a cross-validated $\sqrt{WRSS}$ of (mean [min, max]) 0.95 [0.91, 0.98] years and the \textit{Cox} model (cross-validated $\sqrt{WRSS}$ = 0.95 [0.90, 0.98] years) perform almost equally well. The CART random forest shows the worst performance (cross-validated $\sqrt{WRSS}$ = 1.00 [0.93, 1.03] years). The average treatment effects~$\overline{\hat{\Delta}}(\tau)$ estimated by the two random forest approaches are close to zero (CART random forest: $-$0.14 [$-$2.88, 2.71] days, conditional random forest: 0.38 [$-$0.56, 1.71] days), indicating no advantage of any of the two groups. This supports the results found by de Gregorio et al. \cite{GregorioHaeberleFaschingetal.2020}, who concluded that the interventional treatment does not improve survival in patients with high-risk early breast cancer. In contrast, the \textit{Cox} model indicated a slight advantage of the control group (average treatment effect = 4.35 [$-$0.45, 11.65] days) while the \textit{GEE} model indicated a slight advantage of the interventional group (average treatment effect = $-$3.19 [$-$10.30, 6.03] days). Note, however, that the treatment difference is measured in days, so none of the obtained differences can be considered clinically relevant.

Figure \ref{fig:Results_SUCCESS_A} (B) visualizes the RMST values at $\tau = 5$ years (estimated by the conditional random forest on the complete cohort) in patient groups defined by molecular tumor subtypes \cite{Perou2000}. More specifically, \textit{HER2 positive} patients are characterized by \textit{HER2} positive tumors (regardless of \textit{ER} status, \textit{PR} status and \textit{grade}). \textit{HER2} negative tumors are further classified into \mbox{\textit{luminal A-like}} tumors (\textit{HER2} negative, \textit{ER} and/or \textit{PR} status positive, \textit{grade} G1 or G2), \textit{luminal B-like} tumors (\textit{HER2} negative, \textit{ER} and/or \textit{PR} status positive, \textit{grade} G3), and \textit{triple-negative} tumors (\textit{HER2}, \textit{ER} and \textit{PR} status negative and any~\textit{grade}, see Table~\ref{tab:subgroups_SUCCESS_A}). According to Figure~\ref{fig:Results_SUCCESS_A} (B), the high-risk \textit{triple-negative} group has the lowest estimated RMST values (\mbox{mean (standard deviation): 4.47 (0.28) years}), which is consistent with findings in the literature \cite{vonMinckwitz}. Additionally, when comparing the \textit{luminal A-like} and \textit{luminal B-like} subgroups, there appears to be a slight advantage (corresponding to higher estimated RMST values) of patients with tumor \textit{grade} G1 (\textit{luminal A-like}, 4.78 (0.09) years) compared to those with tumor \textit{grade} G2 or G3 (\textit{luminal B-like}, 4.68 (0.11) years). Again, this result is in line with previous findings in the literature \cite{vonMinckwitz}. The comparison of \textit{luminal A-like}, \textit{luminal B-like} and \textit{triple-negative} confirms the ability of the conditional random forest to identify interactions between hormone receptor status and grade. 

As illustrated in Figure \ref{fig:Results_SUCCESS_A} (C), the $\mbox{PFI}_j$ values obtained from the conditional random forest method identify \textit{nodal status} as the most influential covariate in the estimation of the RMST. The strong influence of \textit{nodal status} on DFS has previously been reported by Senkus et al. \cite{Senkus_2015}. Other important covariates (in terms of $\mbox{PFI}_j$) were \textit{ER}, \textit{stage}, \textit{PR}, \textit{grade}, \textit{HER2} and \textit{BMI}. Notably, all other covariates appear to have negligible importance in estimating RMST values by the conditional random forest method, including \textit{treatment}. Again, this result is in line with the findings of de Gregorio et al. \cite{GregorioHaeberleFaschingetal.2020}. Similar trends in $\mbox{PFI}_j$ were observed for the CART random forest method. 

The local Shapley values in Figure \ref{fig:Results_SUCCESS_A} (D) are also in line with previous findings in the literature \cite{Senkus_2015, vonMinckwitz, goldhirsch} and with the $\mbox{PFI}_j$ values in Figure \ref{fig:Results_SUCCESS_A} (C). As seen from Figure \ref{fig:Results_SUCCESS_A} (D) lymph node positive patients (\textit{nodal status} = pN+) exhibit higher risk of recurrence or death, reflected by negative local Shapley values for these patients. Furthermore, the high Shapley values for \textit{ER} confirm the importance of this covariate in adjuvant hormonal and chemotherapy. The survival advantage of \textit{ER} positive patients \cite{goldhirsch} is reflected by positive Shapley values for this group. Conversely, negative Shapley values are observed for \textit{ER} negative, \textit{PR} negative, and \textit{HER2} negative patients, which is consistent with lower estimated RMST values for the \textit{triple-negative} group in Figure \ref{fig:Results_SUCCESS_A} (B) \cite{vonMinckwitz}. Likewise, the difference in estimated RMST values between \textit{luminal A-like} and \textit{luminal B-like} patients is confirmed by the respective local Shapley values: Patients with \textit{grade} G1 and G2 have a positive contribution to the estimated RMST values, while patients with \textit{grade} G3 contribute negatively. Furthermore, the local Shapley values accurately reflect the hierarchy of tumor stages: Tumor stage pT1 (best prognosis for DFS) has a positive contribution to the estimated RMST, whereas tumor stages pT2 to pT4 have increasingly negative contributions. Shapley values for \textit{treatment} spread around 0 for both groups, suggesting neither a positive nor a negative contribution of the treatment group to the estimated RMST. Again, this result is consistent with the findings in de Gregorio et al.~\cite{GregorioHaeberleFaschingetal.2020}.

\section{Discussion}
\label{Discussion}
During the past years, the restricted mean survival time has become an increasingly popular measure for summarizing individual event times in medical studies. Compared to other established measures like the hazard ratio, the RMST is derived from survival probabilities measured at the untransformed risk scale, thereby avoiding interpretability and collapsibility issues in the comparison of interventional groups \cite{Hernan2010, Didelez2022}. As a consequence, the RMST is considered a valid survival estimand for the causal interpretation of treatment contrasts in clinical and observational trials \cite{Ni_2021, Chen_2001}.

In this work, we proposed the pseudo-value random forest (PVRF) method, which is a non-parametric approach for the quantification of treatment effects by group-specific RMST values. Instead of estimating RMST values from (semi-)parametric models like Cox or AFT regression, the PVRF method combines unconditional pseudo-value RMST estimation with the subsequent fitting of a random forest. Of note, both components of our method (i.e., pseudo-values and random forests) require minimal assumptions on the data-generating process. While unconditional pseudo-values are based on non-parametric Kaplan-Meier estimates, random forest regression is a model-free algorithm allowing for variable selection and requiring no prior assumptions on the structure of the covariate effects. As a result, the PVRF method is particularly suited for incorporating subgroup characteristics, non-linearities, and higher-order interactions affecting individual RMST values. In non-randomized studies, this approach is particularly useful when treatment effects need to be corrected for higher-dimensional sets of confounders, allowing for the estimation of causal contrasts via g-computation (see Section \ref{RMST_difference}). Furthermore, our method enables model-free comparisons of treatment and control group(s) in randomized trials. Regarding the latter, we demonstrated that PVRF is able to capture time-dependent treatment effects in a data-driven way (see Section \ref{Experiments}, where PVRF performed better than (semi-)parametric approaches in the scenarios with crossing survival curves). Methods to adjust RMST estimation for covariate-dependent censoring have been studied in Rong et al.~\cite{Rong_2022}.
In our numerical studies, we observed that the conditional random forest (correcting for a possible selection bias towards covariates with many possible splits) showed a better performance in terms of RMSE than the traditional CART random forest approach. This finding was particularly evident in the estimation of treatment contrasts, where conditional random forests outperformed CART in all scenarios. We therefore recommend to prefer conditional random forests over CART random forests when the aim is to estimate treatment effects from data with heterogeneous covariate types. 
Apart from the estimation of treatment contrasts, the PVRF method can also be used for the {\em prediction} of individual survival times from patient characteristics. In this context, individual RMST differences (keeping all patient characteristics except the treatment variable fixed) can be interpreted as the predicted additional lifetime gained through a specific treatment. Due to the complex structure of random forest predictions (being mainly responsible for the aforementioned flexibility of PVRF), it is usually difficult to quantify the exact strengths and shapes of the covariate effects on the prediction. However, recent advances in interpretable machine learning allow for addressing this issue, as demonstrated by the use of local Shapley values in our application (Section \ref{Application}).
An important topic for future research is the development of hypothesis tests and confidence intervals for PVRF-based RMST differences. Previous research in this field \cite{Royston2013, Tian_2018, Huang_Kuan_2018} has mainly focused on hypothesis tests for RMST differences derived by group-wise integration of Kaplan-Meier curves (not incorporating additional covariates). Similarly, Tian et al.~\cite{Tian_2018} compared RMST-based tests to HR-based tests in the context of randomized clinical trials, demonstrating that RMST-based tests outperformed their HR-based counterparts in scenarios where the PH assumption is violated. It would be interesting to conduct analogous studies for pseudo-value-based tests of RMST differences, which, to the best of our knowledge, have not yet been explored thus far. Confidence intervals for treatment contrasts could, e.g., be constructed using bootstrap methods, along the lines of Hern\'{a}n \& Robins \cite{hernanRobins2020}, Chapter~13. 

A general issue in the estimation of RMST values is the choice of a suitable time horizon $\tau$. While choosing a small value of $\tau$ may discard a large proportion of the data and will therefore result in a potential loss of information, estimation of the RMST may no longer be possible if $\tau$ becomes too large \cite{Tian_2020}. General recommendations on the choice of $\tau$, have, for instance, been made by Tian et al.~\cite{Tian_2020}: Before data collection (for instance, in the course of planning a clinical trial), it is advisable to pre-select~$\tau$ based on clinical and feasibility considerations. If pre-selection of $\tau$ is not possible (e.g.\@ when the analysis is conceived after data collection), Tian et al.\@ suggest to explore a data-dependent time window for $\tau$ and to select the time horizon based on the empirical behavior of the observed times in this window (e.g.\@ by computing quantiles of $\tilde{T}$, as done in our simulations). Alternatively, the RMST could be modeled as a function of $\tau$, as suggested by Zhong \& Schaubel \cite{Zhong_Schaubel_2022}.
We finally note that pseudo-value-based RMST modeling is not restricted to the use of random forest regression. In this work, we focused on random forests because this method is considered to be ``among the best `off-the-shelf' supervised learning methods that are available'' \cite{Coleman_2020}. In particular, random forests are known to perform well on medium-sized data (as often encountered in medical applications), with several efficient software implementations being available \cite{ranger_R_package}. However, it is of course possible to extend our approach to other statistical modeling or machine learning techniques, e.g.\@ to gradient boosting \cite{Schenk_2024} or deep neural networks \cite{Hu_2021, Zhao_2021}. 

\section*{Acknowledgments}
We thank Dr. Lothar Häberle (Department of Gynecology, Obstetrics and Mammology, University Hospital Erlangen, Germany) for supporting us with the analysis of the SUCCESS-A study data.

\bibliography{bibl}

\newpage

\section*{Appendix}

\section*{A \ \ Simulation Study}
\label{supplement:sim_study}
\vspace*{12pt}

\subsection*{A.1 \ \ Covariance matrix of the continuous covariates\label{supplement:covariance_matrix}}

\setcounter{table}{0}
\renewcommand{\thetable}{A\arabic{table}}

\begin{table}[h]
\centering
\begin{tblr}{>{\centering}p{1.4cm}|>{\centering}p{2.5cm}>{\centering}p{1.5cm}>{\centering}p{1.5cm}>{\centering}p{1.5cm}>{\centering}p{1.5cm}}
& $X^{(1)}$ & $X^{(2)}$ & $X^{(3)}$ & $X^{(4)}$ & $X^{(5)}$ \\
\hline 
$X^{(1)}$ & 1.00 & -0.08 & -0.47 & 0.73 & -0.44\\
$X^{(2)}$ & -0.08 & 1.00 & 0.85 & -0.05 & -0.31\\
$X^{(3)}$ & -0.47 & 0.85 & 1.00 & -0.38 & -0.33\\
$X^{(4)}$ & 0.73 & -0.05 & -0.38 & 1.00 & -0.37 \\
$X^{(5)}$ & -0.44 & -0.31 & -0.33 & -0.37 & 1.00\\
\hline 
\end{tblr}
\caption{Covariance matrix of the continuous covariates $X^{(j)}$, $\,j = 1, \ldots, 5$.}
\label{tab:covariance_matrix}
\end{table}

\newpage

\subsection*{A.2 \ \ Values of the time horizon $\tau$ \label{supplement:tau_values}}

\begin{table}[h]
\centering
\begin{tblr}{>{\centering}p{1.4cm}>{\centering}p{2.5cm}>{\centering}p{1.5cm}>{\centering}p{1.5cm}>{\centering}p{1.5cm}>{\centering}p{1.5cm}>{\centering}p{1.5cm}}
\SetCell[r=2]{} \textbf{Scenario} & \textbf{Censoring} & 
\SetCell[r=2]{} \boldmath{$q_{50\%}$} &
\SetCell[r=2]{} \boldmath{$q_{60\%}$} & 
\SetCell[r=2]{} \boldmath{$q_{70\%}$} & 
\SetCell[r=2]{} \boldmath{$q_{80\%}$} & 
\SetCell[r=2]{} \boldmath{$q_{90\%}$} \\
& \textbf{proportion} & & & & & \\
\hline 
\SetCell[r=3]{} 1 & $25\%$ & 1.72 & 2.81 & 4.64 & 8.08 & 16.95\\
& $50\%$ & 1.09 & 1.56 & 2.23 & 3.28 & 5.24\\
& $75\%$ & 0.40 & 0.55 & 0.74 & 1.02 & 1.52\\
\hline
\SetCell[r=3]{} 2 & $25\%$ & 1.37 & 2.38 & 4.26 & 8.19 & 19.64 \\
& $50\%$ & 0.78 & 1.15 & 1.68 & 2.51 & 4.07\\
& $75\%$ & 0.23 & 0.32 & 0.44 & 0.62 & 0.92\\
\hline
\SetCell[r=3]{} 3 & $25\%$ & 1.41 & 1.83 & 2.05 & 2.23 & 2.74\\
& $50\%$ & 0.64 & 0.88 & 1.07 & 1.22 & 1.65\\
& $75\%$ & 0.31 & 0.42 & 0.55 & 0.64 & 0.86\\
\hline
\SetCell[r=3]{} 4 & $25\%$ & 1.22 & 1.67 & 2.03 & 2.19 & 2.77\\
& $50\%$ & 0.52 & 0.74 & 1.01 & 1.14 & 1.57\\
& $75\%$ & 0.20 & 0.27 & 0.36 & 0.43 & 0.60\\
\hline 
\end{tblr}
\caption{Time horizons $\tau$ used in the simulation study. Note that $q_{70\%}$ is approximately equal to the transition time $t_0$ in Scenarios 3 and 4.}
\label{tab:tau_values}
\end{table}

\newpage

\section*{B \ \ Specification and implementation of the \\
\phantom{B} \ \ methods}%
\label{supplement:Implementation}
\vspace*{12pt}

The simulation study and the application were carried out in R, version~4.1.2 \cite{R_referenz}. Data for the simulation study were generated using the R package \texttt{simstudy}, version~0.7.1 \cite{simstudy_R_package}. Pseudo values for the RMST, as defined in Equation \eqref{eq:PV}, were calculated using the \texttt{pseudomean} function of the R package \texttt{pseudo}, version 1.4.3 \cite{pseudo_package}. 

The CART random forest was implemented using the R package \texttt{ranger}, \linebreak version~0.15.1 \cite{ranger_R_package}. The number of trees was set to 500. Data for tree building was sampled without replacement from the complete data using a sampling fraction of 0.632. The number of candidate split variables in each node (``mtry'') was tuned using five-fold cross validation. In each tree, the minimum number of observations required to perform an additional split was set to 5. There were no restrictions on the tree depth and the minimum number of observations in the leaf nodes. 

The conditional random forest was implemented using \texttt{cforest} function of the R package \texttt{partykit}, version 1.2.20 \cite{partykitPackage}. The number of trees was set to 500. By default, \texttt{cforest} implements sampling without replacement from the complete data, using a sampling fraction of 0.632. The number of candidate split variables in each node was tuned using five-fold cross validation. In each tree, a minimum of 20 observations was required to perform a split, and each leaf node was required to contain a minimum of 7 observations. There was no restriction on the depth of the trees. 

The \textit{Lognormal}, \textit{Cox}, and \textit{Reference} methods were implemented using the R package \texttt{survival}, version 3.5.7 \cite{survival_package}. \textit{GEE} and \textit{GEE (log)} were implemented using the R package \texttt{geepack}, version 1.3.9 \cite{geepack_package}. 

Permutation feature importance values were calculated as
\begin{align}
\label{eq:PFI}
\mbox{PFI}^k_j \, = \, \frac{\mbox{WRSS}_k^{j*}}{\mbox{WRSS}_K} \, ,
\end{align}
where $\mbox{WRSS}_k$ denotes the WRSS calculated from the unpermuted data in the $k$-th test fold ($k=1,\ldots , 5$) and $\mbox{WRSS}_k^{j*}$ denotes the respective WRSS calculated from data with randomly permuted values of the $j$-th covariate. Local Shapley values were calculated using the R package \texttt{iml}, version 0.11.2 \cite{iml_R_package}. 

The R-code for the simulation study is available at\\ {\tt https://www.imbie.uni-bonn.de/cloud/index.php/s/6gmJQmayFAMJZHk}.

\newpage

\section*{C \ \ Simulation results for censoring proportions \\
\phantom{C} \ \ $25\%$ and $75\%$}
\label{supplement:results_simulation}
\vspace*{12pt}

\setcounter{figure}{0}
\renewcommand{\thefigure}{C\arabic{figure}}

\begin{figure}[h]
\centering
\includegraphics[width=\textwidth]{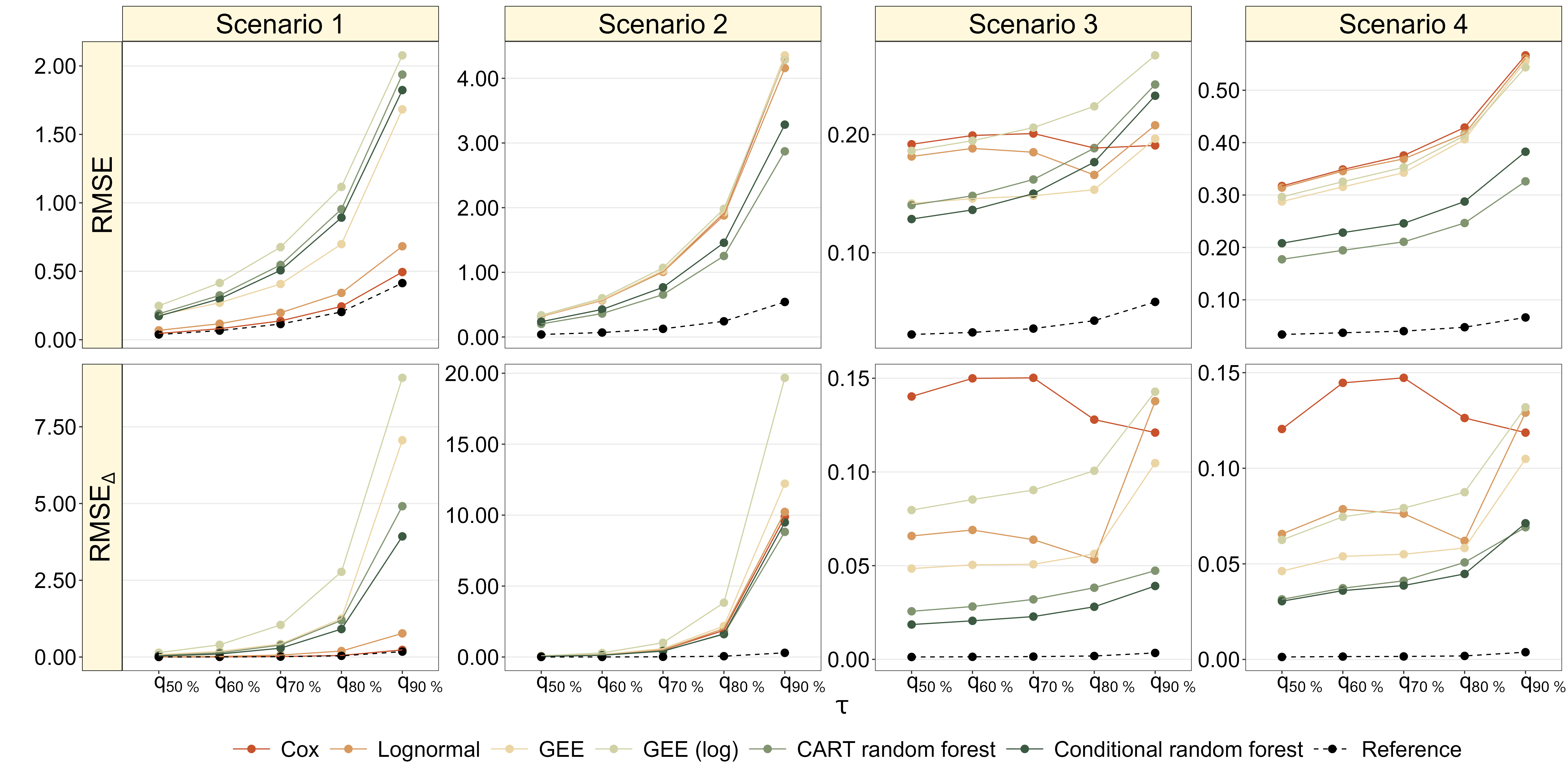}
\caption{Results of the simulation study (25\% censoring). The upper panels present the average RMSE \eqref{eq:MSE}, as obtained from the RMST estimates at different values of~$\tau$. The lower panels present the average RMSE for the treatment effect~\eqref{eq:MSE_delta}. The green colored dots refer to the PVRF method (CART random forest and conditional random forest). The dashed black lines refer to the correctly specified Cox proportional hazards model (\textit{Reference}). } 
\label{fig:Results_Simulations_25}
\end{figure}

\newpage 

\begin{figure}[h]
\centering
\includegraphics[width=\textwidth]{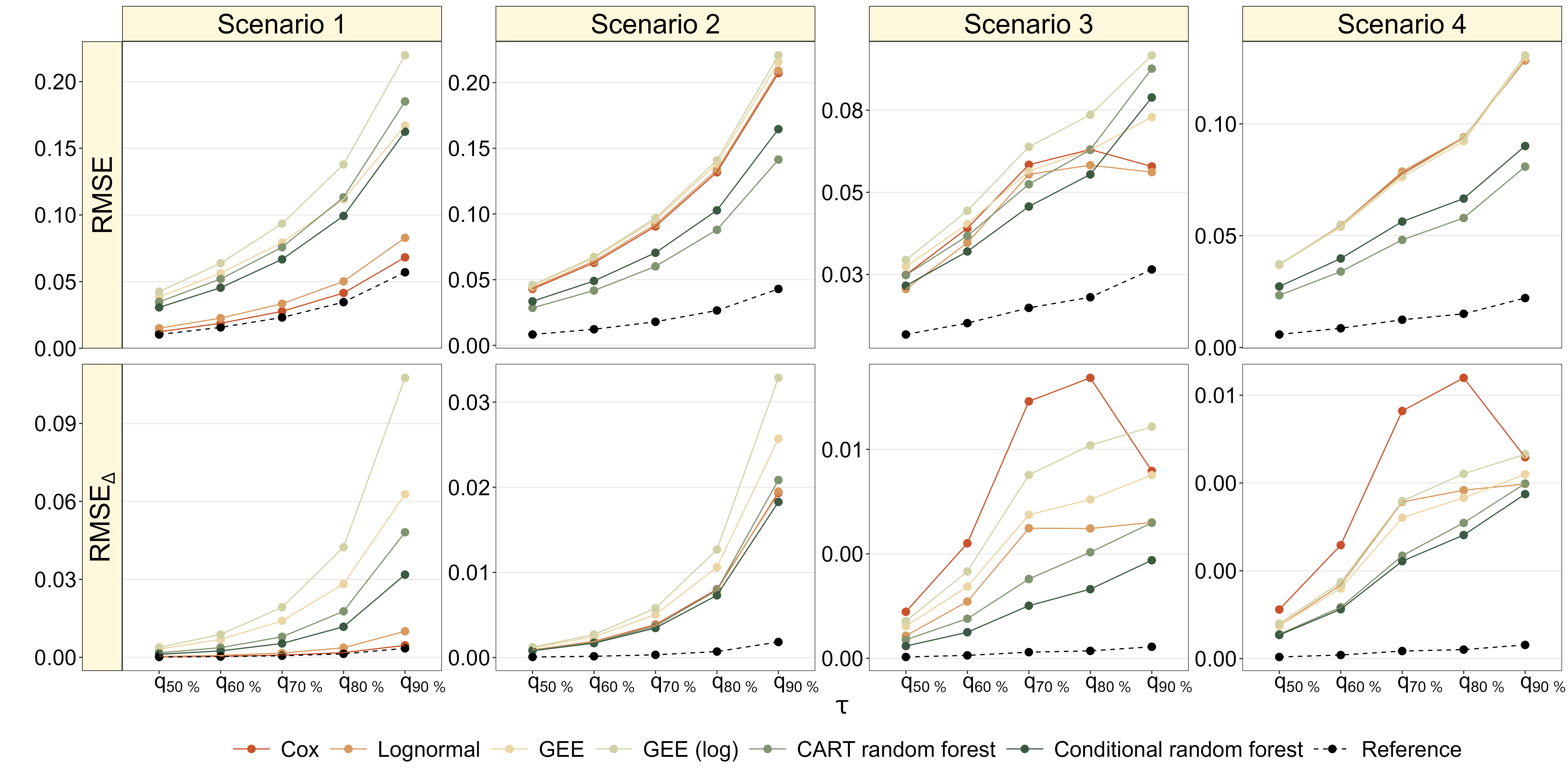}
\caption{Results of the simulation study (75\% censoring). The upper panels present the average RMSE \eqref{eq:MSE}, as obtained from the RMST estimates at different values of~$\tau$. The lower panels present the average RMSE for the treatment effect~\eqref{eq:MSE_delta}. The green colored dots refer to the PVRF method (CART random forest and conditional random forest). The dashed black lines refer to the correctly specified Cox proportional hazards model (\textit{Reference}). } ~
\label{fig:Results_Simulations_75}
\end{figure}

\newpage

\section*{D \ \ Application}
\label{supplement:application}
\subsection*{D.1 \ \ Patient characteristics of the SUCCESS-A study data\label{supplement:patients}}

\setcounter{table}{0}
\renewcommand{\thetable}{D\arabic{table}}

\begin{table}[h!]
\centering
\scriptsize
\begin{tblr}{>{}m{4.5cm}>{}m{4cm}>{}m{5.5cm}}
\textbf{Characteristic} & 
\textbf{} & 
\textbf{Patients (N = 3,754)} \\
\hline
\textbf{age (years)} & 
Mean (SD) & 53.5 (10.5) \\ 
& Median [Min, Max] & 53.0 [21.0, 86.0] \\ 
\hline
\textbf{BMI (\boldmath{$kg/m^2$})} & 
Mean (SD) & 26.3 (5.03) \\ 
& Median [Min, Max] & 25.4 [15.4, 53.4] \\ 
\hline
\textbf{tumor stage} & 
pT1 & 1552 (41.3\%) \\ 
& pT2 & 1929 (51.4\%) \\ 
& pT3 & 198 (5.3\%) \\ 
& pT4 & 52 (1.4\%) \\ 
& Missing & 23 (0.6\%) \\ 
\hline
\textbf{tumor grade} & 
G1 & 176 (4.7\%) \\ 
& G2 & 1783 (47.5\%) \\ 
& G3 & 1773 (47.2\%) \\ 
& Missing & 22 (0.6\%) \\ 
\hline
\textbf{lymph node status} & pN+ & 2452 (65.3\%) \\ 
& pN0 & 1273 (33.9\%) \\ 
& Missing & 29 (0.8\%) \\ 
\hline
\textbf{tumor type} & ductal & 3060 (81.5\%) \\ 
& lobular & 419 (11.2\%) \\ 
& other & 253 (6.7\%) \\ 
& Missing & 22 (0.6\%) \\ 
\hline
\textbf{estrogen receptor status} & ER- & 1252 (33.4\%) \\ 
& ER+ & 2481 (66.1\%) \\ 
& Missing & 21 (0.6\%) \\ 
\hline
\textbf{progesterone receptor status} & PR- & 1525 (40.6\%) \\ 
& PR+ & 2205 (58.7\%) \\ 
& Missing & 24 (0.6\%) \\ 
\hline
\textbf{HER2} & HER2- & 2787 (74.2\%) \\ 
& HER2+ & 883 (23.5\%) \\ 
& Missing & 84 (2.2\%) \\ 
\hline
\textbf{menopausal status} & pre & 1565 (41.7\%) \\ 
& post & 2189 (58.3\%) \\ 
\hline
\textbf{treatment group} & control & 1898 (50.6\%) \\ 
& interventional & 1856 (49.4\%) \\ 
\hline 
\end{tblr}
\caption{Descriptive summary of the variables in the SUCCESS-A study data.}
\label{tab:supplement_SUCCESS_A}
\end{table}

\newpage

\subsection*{D.2 \ \ Definition of molecular tumor subtypes \label{supplement:classification}}

\begin{table}[h!]
\centering
\begin{tblr}{>{\centering}m{4cm} >{\centering}m{1.5cm} >{\centering}m{4cm} >{\centering}m{2.3cm}}
\textbf{Subgroup} & \textbf{HER2} & \textbf{ER/PR} & \textbf{grade} \\
\hline 
\textbf{HER2 positive} & HER2+ & any & any\\
\textbf{luminal A-like} & HER2$-$ & ER+ and/or PR+ & G1 or G2\\
\textbf{luminal B-like} & HER2$-$ & ER+ and/or PR+ & G3\\
\textbf{triple-negative} & HER2$-$ & ER$-$ and PR$-$ & any\\
\hline 
\end{tblr}
\caption{Definition of molecular tumor subtypes.}
\label{tab:subgroups_SUCCESS_A}
\end{table}

\newpage

\subsection*{D.3 \ \ Supplementary results\label{supplement:application_results}}

\setcounter{figure}{0}
\renewcommand{\thefigure}{D\arabic{figure}}

\begin{figure}[h]
\centering
\includegraphics[width=0.87\textwidth]{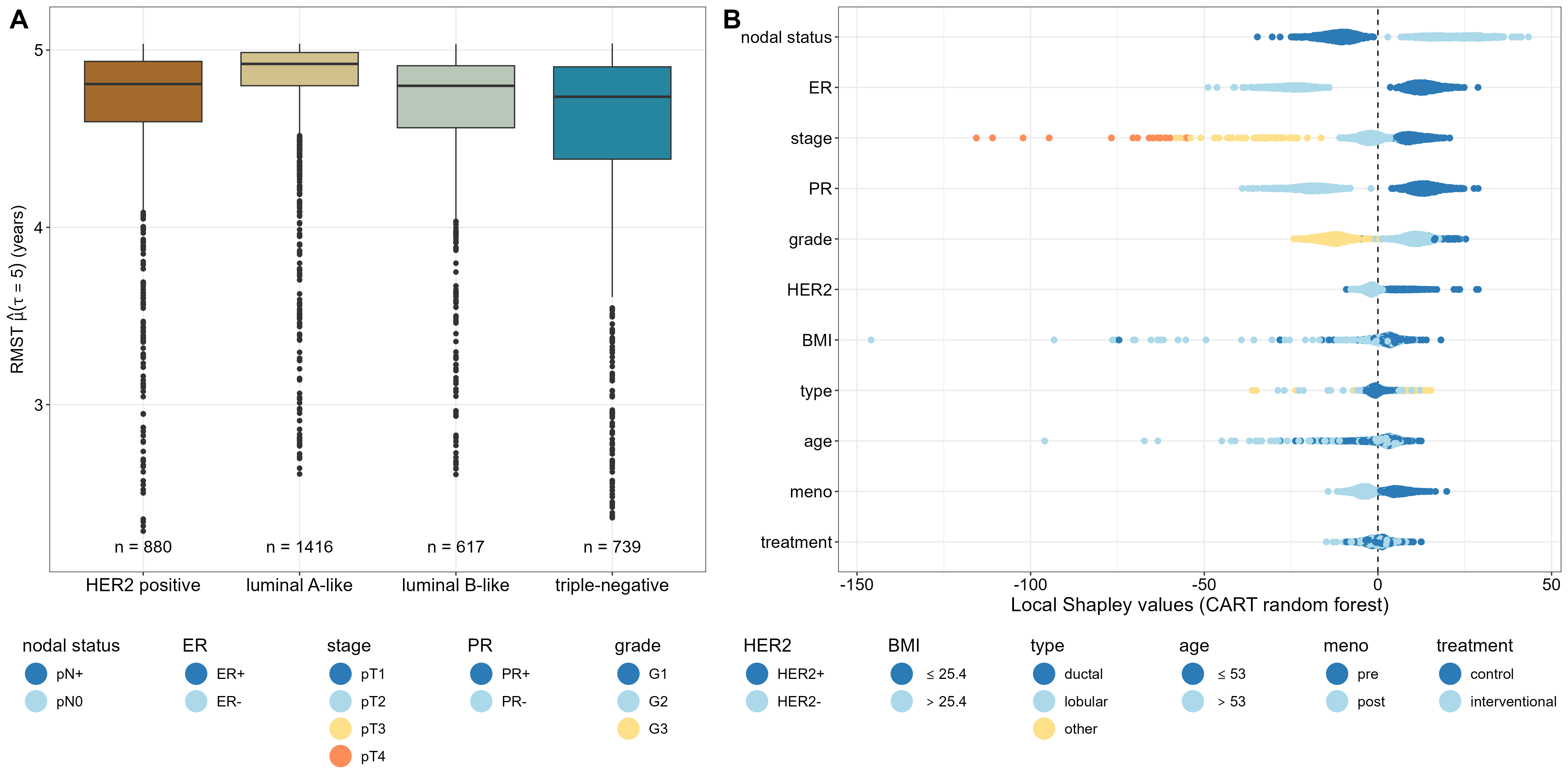}
\caption{Analysis of DFS in the SUCCESS-A study. Panel (A) shows the estimated RMST values in patient groups defined by molecular tumor subtypes, as obtained from the CART random forest (see Table \ref{tab:subgroups_SUCCESS_A}). Panel (B) presents local Shapley values for each covariate, as obtained by evaluating the CART random forest estimates in 1000 randomly selected patients. Each dot corresponds to one patient. The color codings used in Panel (B) are presented in the bottom row of the figure.} 
\label{fig:SUCCESS_A_CART}
\end{figure}

\end{document}